\documentclass[aps,prmaterials,bibnotes,a4paper,10pt,twocolumn,showpacs,floatfix,superscriptaddress,amsmath,preprintnumbers,citeautoscript,longbibliography]{revtex4-1}

\usepackage[utf8]{inputenc} 

\usepackage{float}
\usepackage{dcolumn,graphicx,color,booktabs,microtype,afterpage}

\usepackage[charter,greekuppercase=italicized]{mathdesign}

\usepackage{booktabs}
\usepackage{longtable}
\usepackage{sidecap}

\renewcommand{\tablename}{Table}
\makeatletter\renewcommand{\fnum@figure}[1]{\figurename~\thefigure.~}\makeatother
\makeatletter\renewcommand{\fnum@table}[1]{\tablename~\thetable.}\makeatother

\newcount\hh \newcount\mm
\hh=\time \divide\hh by 60
\mm=\hh \multiply\mm by 60 \mm=-\mm
\advance\mm by \time
\def\now{\number\hh:\ifnum\mm<10{}0\fi\number\mm}

\usepackage[colorlinks,plainpages=false,linkcolor=blue,urlcolor=blue,citecolor=blue,pdfpagemode=UseNone,pdfstartview=FitBH]{hyperref}

\newcommand{\tcr}[1]{\textcolor{black}{#1}}

\usepackage{nicefrac}
\newcommand{\CUPZ}{Cu(pz)$_2$(ClO$_4$)$_2$}

\begin{document}

\makeatletter\renewcommand{\ps@plain}{%
\def\@evenhead{\hfill\itshape\rightmark}%
\def\@oddhead{\itshape\leftmark\hfill}%
\renewcommand{\@evenfoot}{\hfill\small{--~\thepage~--}\hfill}%
\renewcommand{\@oddfoot}{\hfill\small{--~\thepage~--}\hfill}%
}\makeatother\pagestyle{plain}

\title{Room-temperature structural phase transition in the quasi-2D spin-\texorpdfstring{$\nicefrac{1}{2}$}{1/2} Heisenberg antiferromagnet 
\texorpdfstring{Cu(pz)$_2$(ClO$_4$)$_2$}{Cu(pz)2(ClO4)2}}

\author{N. Barbero}\email[Corresponding author: \vspace{8pt}]{nbarbero@phys.ethz.ch}
\affiliation{Laboratorium f\"ur Festk\"orperphysik, ETH Z\"urich, CH-8093 Zurich, Switzerland}

\author{M. Medarde}
\affiliation{Laboratory for Multiscale Materials Experiments, Paul Scherrer Institut, CH-5232 Villigen PSI, Switzerland}

\author{T. Shang}
\affiliation{Laboratory for Multiscale Materials Experiments, Paul Scherrer Institut, CH-5232 Villigen PSI, Switzerland}

\author{D. Sheptyakov}
\affiliation{Laboratory for Neutron Scattering and Imaging, Paul Scherrer Institut, CH-5232 Villigen PSI, Switzerland}

\author{C. P. Landee}
\affiliation{Department of Physics, Clark University, Worcester, Massachusetts 01610, USA}

\author{J. Mesot}
\affiliation{Laboratorium f\"ur Festk\"orperphysik, ETH Z\"urich, CH-8093 Zurich, Switzerland}
\affiliation{Paul Scherrer Institut, CH-5232 Villigen PSI, Switzerland}

\author{H.-R. Ott}
\affiliation{Laboratorium f\"ur Festk\"orperphysik, ETH Z\"urich, CH-8093 Zurich, Switzerland}
\affiliation{Paul Scherrer Institut, CH-5232 Villigen PSI, Switzerland}

\author{T. Shiroka}\email[Corresponding author: \vspace{8pt}]{tshiroka@phys.ethz.ch}
\affiliation{Laboratorium f\"ur Festk\"orperphysik, ETH Z\"urich, CH-8093 Zurich, Switzerland}
\affiliation{Paul Scherrer Institut, CH-5232 Villigen PSI, Switzerland}

\begin{abstract}
Cu(pz)$_2$(ClO$_4$)$_2$ (with \textit{pz} denoting pyrazine C$_4$H$_4$N$_2$) 
is a two-dimensional spin-$\nicefrac{1}{2}$ square-lattice antiferromagnet 
with $T_{\mathrm{N}} = 4.24$\,K. Due to a persisting focus on the low-temperature 
magnetic properties, its room\--tem\-pe\-rature structural and physical 
properties caught no attention up to now. Here we report a study of the 
structural features of Cu(pz)$_2$(ClO$_4$)$_2$ in the paramagnetic phase, 
up to 330\,K. By employing magnetization, specific heat, $^{35}$Cl nuclear 
magnetic resonance, and neutron diffraction measurements, we provide 
evidence of a second-order phase transition at $T^{\star} = 294$\,K, not 
reported before. 
The absence of a magnetic ordering across $T^{\star}$ in the magnetization 
data, yet the presence of a sizable anomaly in the specific heat, 
suggest a \tcr{structural order-to-disorder type transition.} 
NMR and neu\-tron-dif\-frac\-tion data corroborate our conjecture, by revealing 
subtle angular distortions of the pyrazine rings and of 
ClO$^-_4$ count\-er\-an\-ion tetrahedra, shown to adopt a configuration 
of higher symmetry above the transition temperature.
\end{abstract}


\maketitle\enlargethispage{3pt}

\section{Introduction\label{sec:intro}}
\begin{figure}[htb]
\centering
  \includegraphics[width=0.45\textwidth]{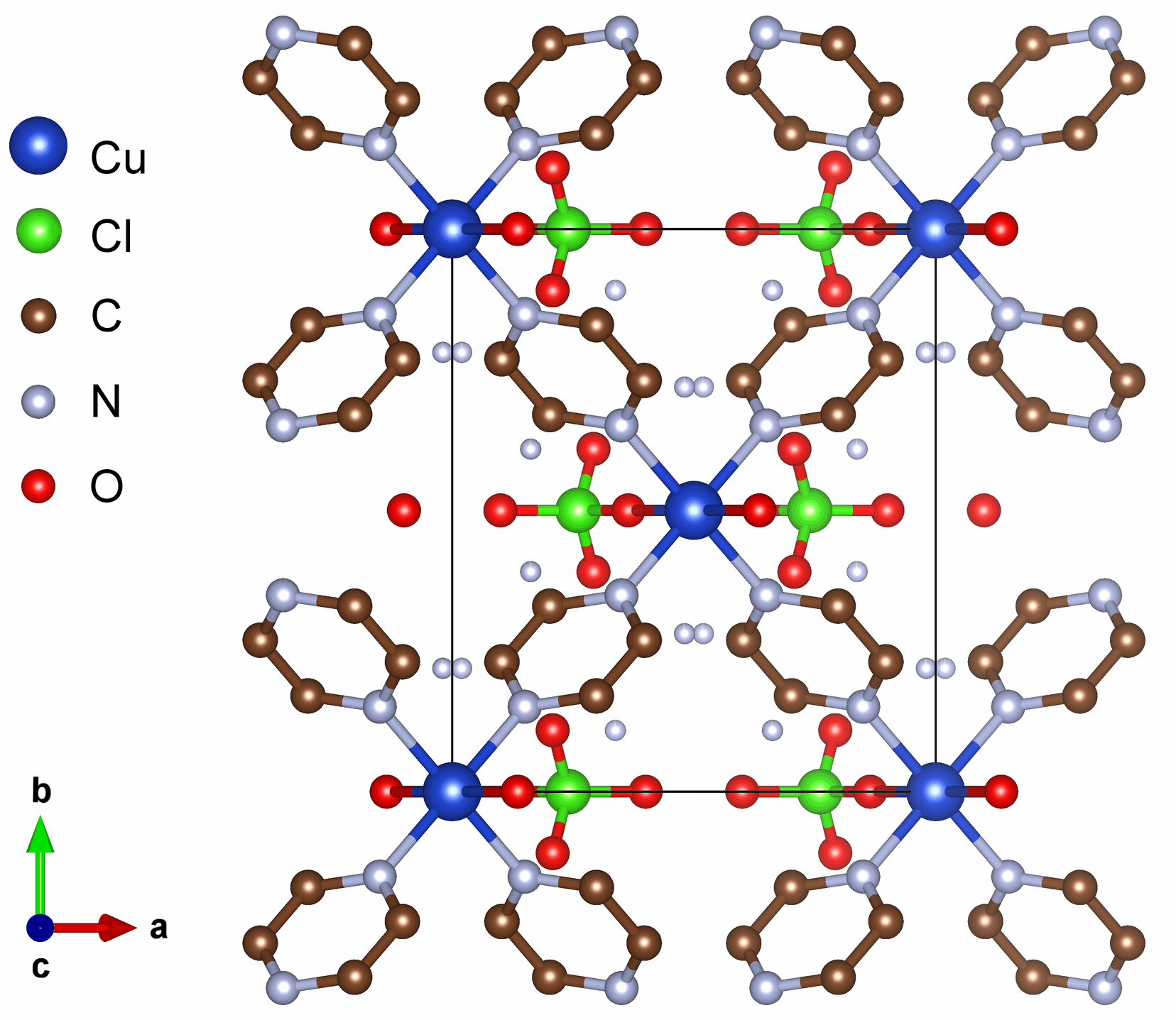}
  \caption{A layer of Cu(pz)$_2$(ClO$_4$)$_2$, whose $C2/m$ structure 
  (space group no.\ 12) was determined from neutron diffraction data at 
  294\,K. Each Cu(II) atom (blue) is linked via pyrazine rings (C = brown, 
  N = gray) to four other copper ions, all lying in the $ab$-plane. 
  The layers are spaced by perchlorate counteranions (Cl = green, 
  O = red), lying along the interlayer $c$-axis direction. As described 
  in the text and in Table~\ref{tab:structure} (see Appendix), the N and 
  C atoms in the pyrazine rings and the O atoms in the perchlorate 
  counteranions have different Wyckoff positions. Each counteranion contains 
  a Cl site at the center and four O atoms at its vertices.
   }
\label{fig:structure}
\end{figure}

\enlargethispage{8pt}
As a notable physical realization of a quasi-2D Heisenberg antiferromagnet, 
\CUPZ\ has been a test case for investigating the competition between long-range 
magnetic order and quantum fluctuations \cite{Darriet1979, Choi2003}.  
Its structure consists of stacks along the $c$-axis of well-isolated 
nearly-square layers of Cu$^{2+}$ ions in the $ab$-plane, rotated by 
45$^\circ$ with respect to the in-plane primitive vectors, 
as shown schematically in Fig.~\ref{fig:structure}. 
Along the $c$-axis, each layer is shifted by half a unit cell along the 
$a$- and $b$-axes. Each Cu$^{2+}$ ion is bridged to its four 
nearest-neighbors (NN) by C$_4$H$_4$N$_2$ pyrazine ligands, which provide 
the intralayer superexchange interaction. Two ClO$_4^{-}$ perchlorate 
counteranions, linked to Cu$^{2+}$ ions via one of the oxygen atoms in 
the O$_4$ tetrahedra, provide a sufficient interlayer separation along 
the $c$-axis, hence implying a negligible interlayer interaction. 
Overall, this results in a Cu$^{2+}$-ion arrangement of nearly-tetragonal 
symmetry \cite{Landee2013}. The 3D antiferromagnetism (AFM) of Cu(pz)$_2$(ClO$_4$)$_2$ 
has been extensively studied by inelastic neutron scattering (INS) \cite{Tsyrulin2010}, 
muon-spin rotation ($\mu$SR) \cite{Lancaster2007}, and nuclear magnetic 
resonance (NMR) \cite{Barbero2016} measurements. 
It has also been shown that applied magnetic fields along the $c$-axis 
strengthen the AFM order by suppressing the quantum fluctuations, hence 
enhancing $T_\mathrm{N}$ above its zero-field value of 4.24\,K \cite{Tsyrulin2010}.
On the other hand, external hydrostatic pressure reduces $T_\mathrm{N}$ 
\cite{Barbero2016}, most likely by enforcing 
1D-type interactions \cite{Mermin1966}, 
as suggested by results of density-functional theory (DFT) calculations 
on similar compounds \cite{Vela2013,Wehinger2016}.

The crystal structure of \CUPZ\ was determined at 163 and 293\,K from 
single-crystal x-ray diffraction data \cite{Woodward2007}, obtaining a better 
refinement with the $C2/c$ space group in the first case
and with the $C2/m$ space group in the latter. It was also observed that the four pyrazine 
moieties form two sets at low temperature, each of them characterized by a different tilting angle with respect to 
the $ab$-plane. The values of these angles are distinct at 163\,K, but they
become identical (65.8$^{\circ}$) at 293\,K. 
Along with x-ray diffraction patterns, Choi et al.\  
\cite{Choi2003} reported also infrared-spectroscopy data, used to track 
the evolution of the vibrational modes as a function of temperature. 
Upon heating, the latter measurements indicated a softening of 
the vibrational modes starting at approximately 180\,K, related to the out-of-plane deformations of 
the pyrazine rings.

In this work, by combining data from NMR, 
specific-heat and neutron-diffraction experiments, we provide clear 
evidence for a \emph{structural phase transition} occurring at 
$T^{\star}$ = 294(1)\,K, not reported before in the literature. 
Our results indicate that the two initially different Cl sites become 
equivalent above $T^{\star}$ and that the symmetry of the individual 
pyrazine rings increases upon heating above $T^{\star}$.

After introducing the experimental details in Sec.~\ref{sec:Experimental_Details}, 
in Sec.~\ref{ssec:squid} and \ref{ssec:heat}, we describe the 
material characterization via magnetization and specific-heat 
measurements. From these data, we identify the onset of the AFM order 
at $T_\mathrm{N}$ and of the structural transition at $T^{\star}$, 
respectively.
The ${}^{35}$Cl NMR results discussed in Sec.~\ref{ssec:nmr} clearly 
show the merging of the two lines upon heating, reflecting an increase 
in structure symmetry and allowing for an evaluation of the isotropic 
hyperfine coupling. Finally, to precisely identify the variations in 
bond lengths and angles, we employed neutron powder diffraction measurements, 
whose results are reported in Sec.~\ref{ssec:neutron}. The combined 
data of our investigations indicate that the second-order structural phase 
transition is accompanied by subtle transformations in both the 
ClO$^-_4$ tetrahedra and in the C$_4$H$_4$N$_2$ pyrazine ligands.

\vspace{5pt}

\section{\label{sec:Experimental_Details}Experimental details}
The Cu(ClO$_4$)$_2$ crystals were synthesized by dissolving 
Cu(ClO$_4$)$_2\cdot$6H$_2$O and pyrazine in water with a drop of dilute 
HClO$_4$(aq), in order to prevent the precipitation of Cu(OH)$_2$ and 
of CuCO$_3$. The solution was then partially covered and left to 
evaporate slowly, with the crystals growing over several weeks. Finally, 
the mixture was filtered, the recovered crystals were washed in 
deionized water at $\sim$ 10$^\circ$C, and dried in air. Since the 
samples are hygroscopic, they were stored in an inert-gas atmosphere 
\cite{Woodward2007}. Single crystals with a typical mass of $\sim$80\,mg 
were aligned with the $c$-axis (\textit{hard} axis) parallel to the 
applied magnetic field, for both the magnetization- and NMR experiments. 
Given the easily identifiable $ab$-planes, delimited by smooth surfaces, 
the visual crystal alignment was achieved with an uncertainty of about 
5$^{\circ}$. For the neutron diffraction measurements, we used 
deuterated powder \tcr{synthesized by following the same protocol, 
but employing deuterated reagents}. 
The magnetization measurements were performed with a commercial magnetic 
property measurement system (MPMS) XL setup from 2 to 310\,K in an 
applied magnetic field of 5\,mT.
The heat-capacity data were collected by means of a Quantum Design 
physical property measurement system (PPMS-9\,T) by using the relaxation 
method, both upon heating and upon cooling, in a temperature range 
between 250 and 330\,K.

The $^{35}$Cl-NMR study comprised lineshape and spin-lattice relaxation 
time measurements in a 7.063-T field, corresponding to a Larmor frequency 
of 29.4664\,MHz for the spin-3/2 $^{35}$Cl quadrupolar nuclei. NMR 
signals were detected by employing a standard spin-echo sequence, 
consisting in ${\pi}/{2}$ and $\pi$ pulses of 2 and 4\,$\mu$s, 
respectively, with recycle delays ranging from 0.6 to 0.2\,s for 
temperatures in the 4--310\,K range. The NMR lineshapes were obtained 
from fast Fourier transforms (FFT) of the echo signals. 
The spin-lattice relaxation times $T_1$ were measured 
via the inversion-recovery method using a  $\pi$-${\pi}/{2}$-$\pi$ pulse sequence.

Neutron diffraction experiments were performed at the HRPT diffractometer 
(high-resolution powder diffractometer for thermal neutrons) at the 
SINQ facility of the Paul Scherrer Institute (PSI) in Villigen, 
Switzerland. Several patterns were recorded between 260 and 330\,K, using 
two different wavelengths ($\lambda$ = 1.494\,\AA\ and 1.886\,\AA) of 
the high-intensity mode (24' primary-beam collimation) \cite{Fischer2000}.
The sample was placed in an 8-mm-diameter vanadium cylinder under helium 
atmosphere and sealed with indium wire. The cylinder was then introduced 
into a cryofurnace whose contribution to the neutron-count background 
was minimized by means of an oscillating radial collimator. All the 
patterns were refined by using the Rietveld refinement FullProf 
suite \cite{Carvajal1993}.
\begin{figure}[htb]
\centering
  \includegraphics[width=0.45\textwidth]{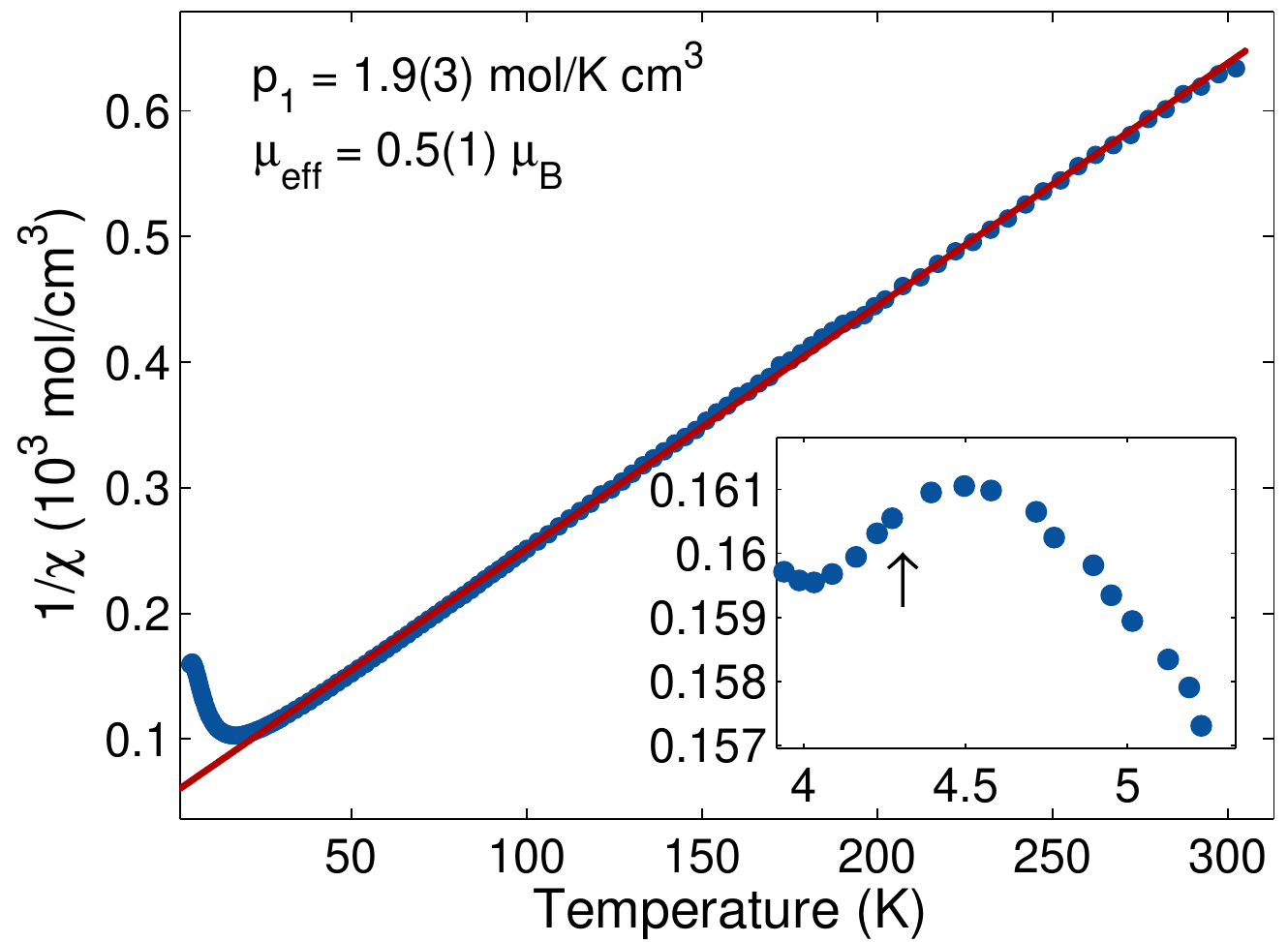}
  \caption{\label{fig:squid}$1/\chi(T)$, as measured at 5\,mT. 
  The linear fit (red line) to the data in the paramagnetic regime 
  ($T > 20$\,K) implies an effective magnetic moment of 0.5(1)\,$\mu_{\mathrm{B}}$, 
  while the negative intercept on the $T$ axis indicates AFM interactions 
  among the Cu$^{2+}$ ions. Inset: low-temperature 1/$\chi(T)$ data. 
  The arrow at $T_{\mathrm{N}} = 4.24$\,K indicates the inflection point, 
  considered as the onset of AFM order \cite{Barbero2016}.}
\end{figure}

\vspace{-5pt}
\section{Results and discussion\label{sec:res}}

\subsection{Magnetization measurements\label{ssec:squid}}
The inverse magnetic susceptibility 1/$\chi(T)$ measured at 5\,mT 
(see Fig.~\ref{fig:squid}) indicates an effective magnetic moment, 
$\mu_\mathrm{eff}$ = 0.5(1)\,$\mu_\mathrm{B}$. Its value, far smaller 
than the corresponding Cu single-ion spin-only magnetic moment 
(1.73\,$\mu_\mathrm{B}$), suggests the presence of strong quantum 
fluctuations. As previously reported \cite{Tsyrulin2010}, an enhancement 
of $\mu_\mathrm{eff}$ in higher magnetic fields may be interpreted as a 
field-induced suppression of the quantum fluctuations. 
Besides the clear anomaly at $T_\mathrm{N} \sim 4.2$\,K, the $\chi_m(T)$ 
response remains paramagnetic up to 300\,K, thus excluding the magnetic 
nature of the transition observed at $T^{\star} = 294$\,K (see below).

\subsection{\label{ssec:heat}Specific-heat measurements}
\begin{figure}[htb]
\centering
  \includegraphics[width=0.45\textwidth]{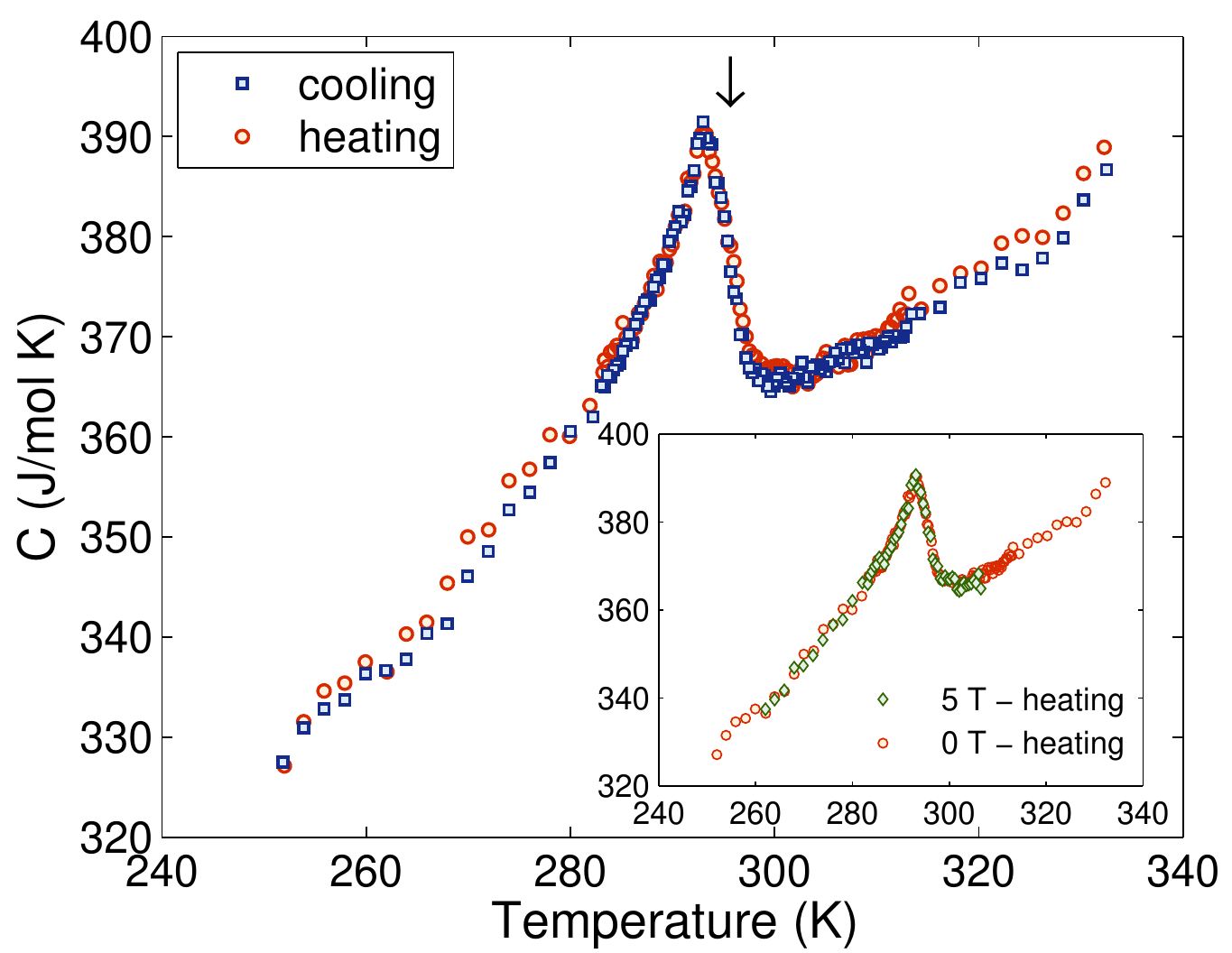}
  \caption{The temperature dependence of the heat capacity measured 
  upon heating and cooling \tcr{showing the lack of hysteresis}. The arrow 
  indicates $T^{\star}$ = 294\,K. Inset: the \tcr{coinciding} 0- and 5-T 
  datasets rule out heat-capacity changes in \tcr{the applied magnetic field range}.}
  \label{fig:spec_heat}
\end{figure}

The heat capacity data vs.\ temperature, collected between 250 and 330\,K 
upon both heating and cooling, are shown in Fig.~\ref{fig:spec_heat}. 
The significant and overlapping (i.e., hysteresis-free) anomalies both 
peak at 293\,K and indicate the existence of a previously unnoticed 
second-order phase transition in this temperature range. The 
independence of the transition from the applied magnetic field 
(of 5\,T in this case, see inset in Fig.~\ref{fig:spec_heat}) and 
the absence of an anomaly in the magnetic susceptibility (see 
Fig.~\ref{fig:squid}), rule out a magnetic origin of the transition 
and suggest the transition at $T^{\star}$ to be of \emph{structural character}. 
From the heat-capacity data, we calculated numerically the entropy 
$\Delta S(T) = \int_{T_0}^T C_p \mathrm{d}(\ln T)$, with $T_0$ 
being a reference temperature. Since across $T^{\star}$ we find 
$\Delta S(T) \sim R$ ($=8.31$\,JK$^{-1}$mol$^{-1}$), this suggest 
an order-to-disorder (i.e., a non displacive) type of structural 
transition (see below).

\subsection{\label{ssec:nmr}Nuclear magnetic resonance measurements}
\begin{figure}[htb]
\centering
  \includegraphics[width=0.45\textwidth]{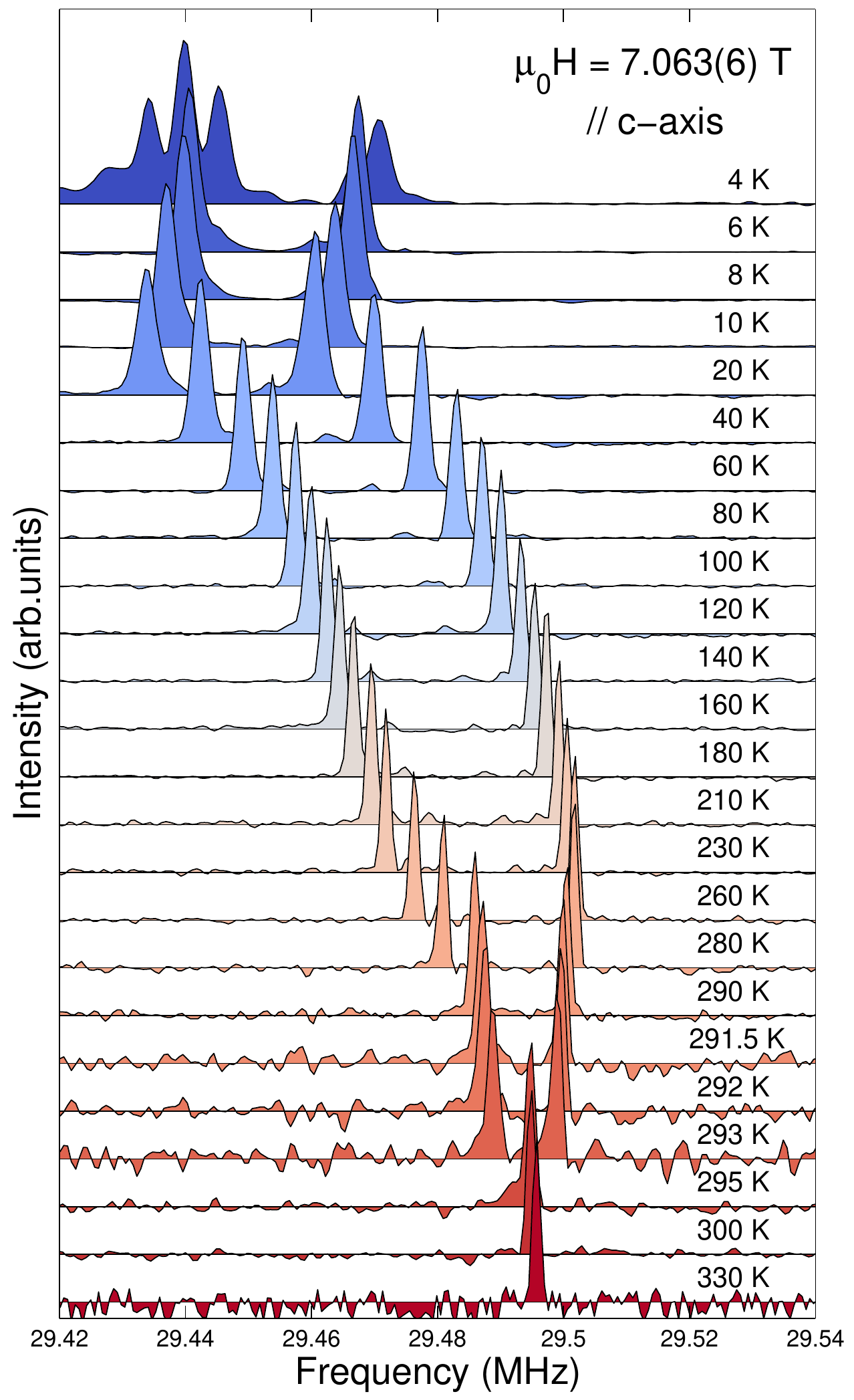}
  \caption{Evolution of the $^{35}$Cl NMR line shape and position with 
  temperature. Note that at room temperature \CUPZ\ exhibits a positive 
  paramagnetic shift of $\sim$0.1\%, with respect to the Larmor frequency 
  (29.466\,MHz in a 7.063-T magnetic field). At $T^{\star} = 294$\,K, the 
  two NMR lines resulting from the two inequivalent chlorine sites merge 
  into a single peak, indicating that the structural transformation of 
  the ClO$_4^{-}$ tetrahedra is complete.}
  \label{fig:lines}
\end{figure}
\begin{figure}[htb]
\centering
  \includegraphics[width=0.45\textwidth]{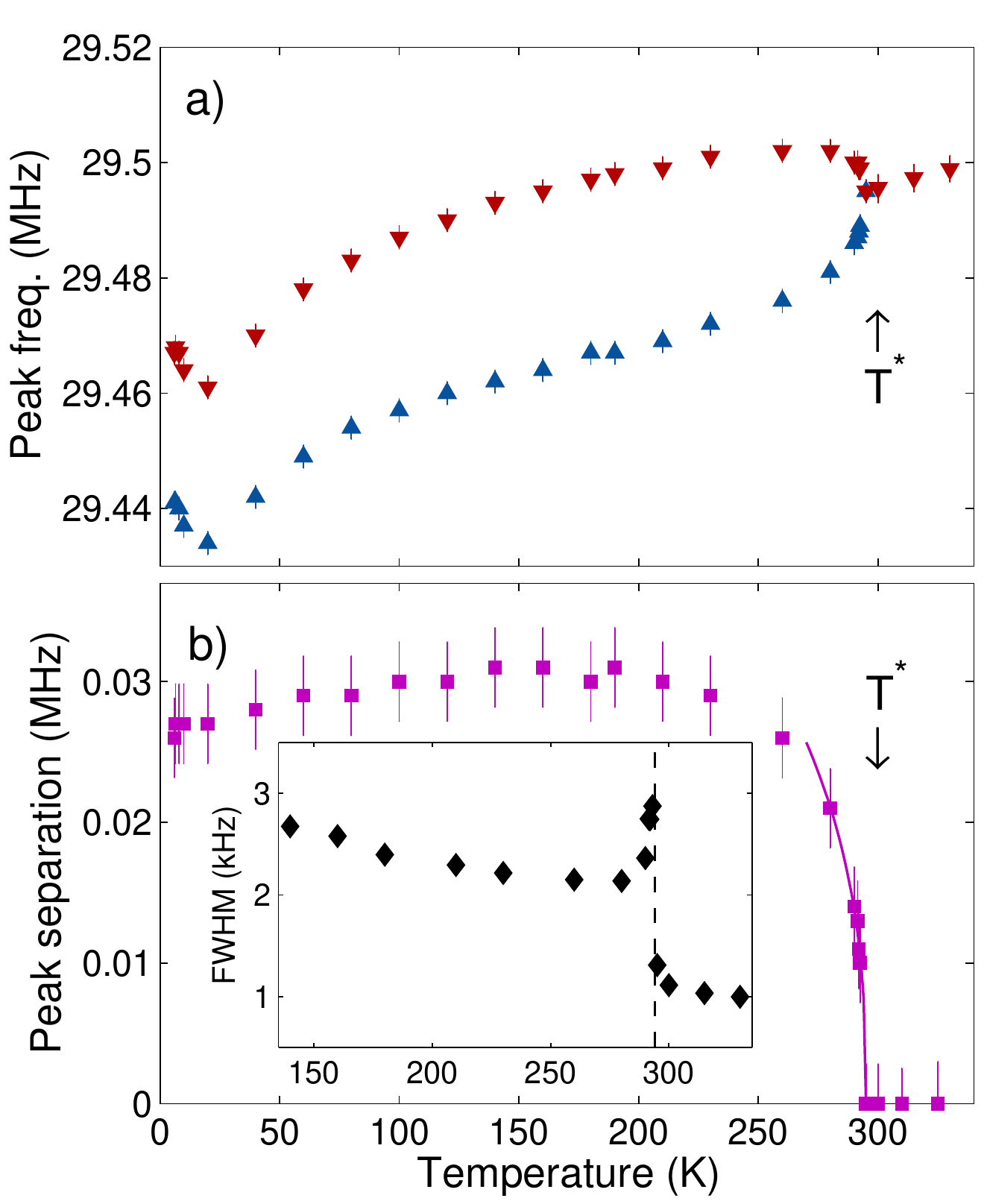}
  \caption{\label{fig:metadata}$^{35}$Cl NMR peak positions vs.\ 
  temperature (a). Despite a general increase in frequency with 
  temperature, the peak separation (b) stays mostly constant 
  (ca.\ $\sim$30\,kHz), to smoothly go to zero at 
  $T^{\star} = 294$\,K (see arrows). The line is a fit to 
  $(1-T/T^{\star})^{\beta}$, with $\beta = 0.45(5)$. Inset: upon heating,   
  the NMR linewidth shows a clear drop at $T^{\star}$ (dashed line) 
  indicative of a transition to a configuration with higher local symmetry.}
\end{figure}

The ${}^{35}$Cl NMR lines measured at different temperatures are shown 
in Fig.~\ref{fig:lines}. The formation of two magnetic sublattices in the 
AFM phase, below 4.24\,K \cite{Barbero2016}, is clearly manifested in 
the line splitting and broadening of the $T=4$\,K dataset.  
Most important to us is the line behavior in the high-temperature regime.
Figure~\ref{fig:metadata} shows the evolution of the peak positions (a) 
and of their separation (b) with temperature. 
Below $T^{\star} = 294$\,K, the two distinct NMR lines correspond to 
two crystallographically inequivalent chlorine sites. Their merging at 
$T^{\star}$ suggests a structural transition of certain structural sub-units 
to a higher symmetry, which reestablishes the site equivalence and 
preserves it upon heating up to at least 330\,K. This transition is 
even better illustrated by the \tcr{peak-to-peak distance} vs.\ 
temperature \tcr{plot} [see Fig.~\ref{fig:metadata}(b)]. The 
\tcr{peak separation} exhibits first a shallow maximum centered 
at 175\,K, then it continuously decreases, to finally vanish at $T^{\star}= 294$\,K, a 
value very close to that of the maximum in the heat-capacity $C(T)$ 
data (293\,K). 

Note that both the Knight shift and the static susceptibility \cite{Barbero2016} 
(i.e., the magnetization data), exhibit a very similar temperature 
dependence. In particular, the susceptibility data were compared to those 
calculated by quantum Monte Carlo simulations and, upon assuming the 
validity of a 2D Heisenberg model on a square lattice (see Fig.~3 in 
Ref.~\onlinecite{Barbero2016}), an excellent agreement was found. 
%
%
\begin{figure}[htb]
\centering
  \includegraphics[width=0.45\textwidth]{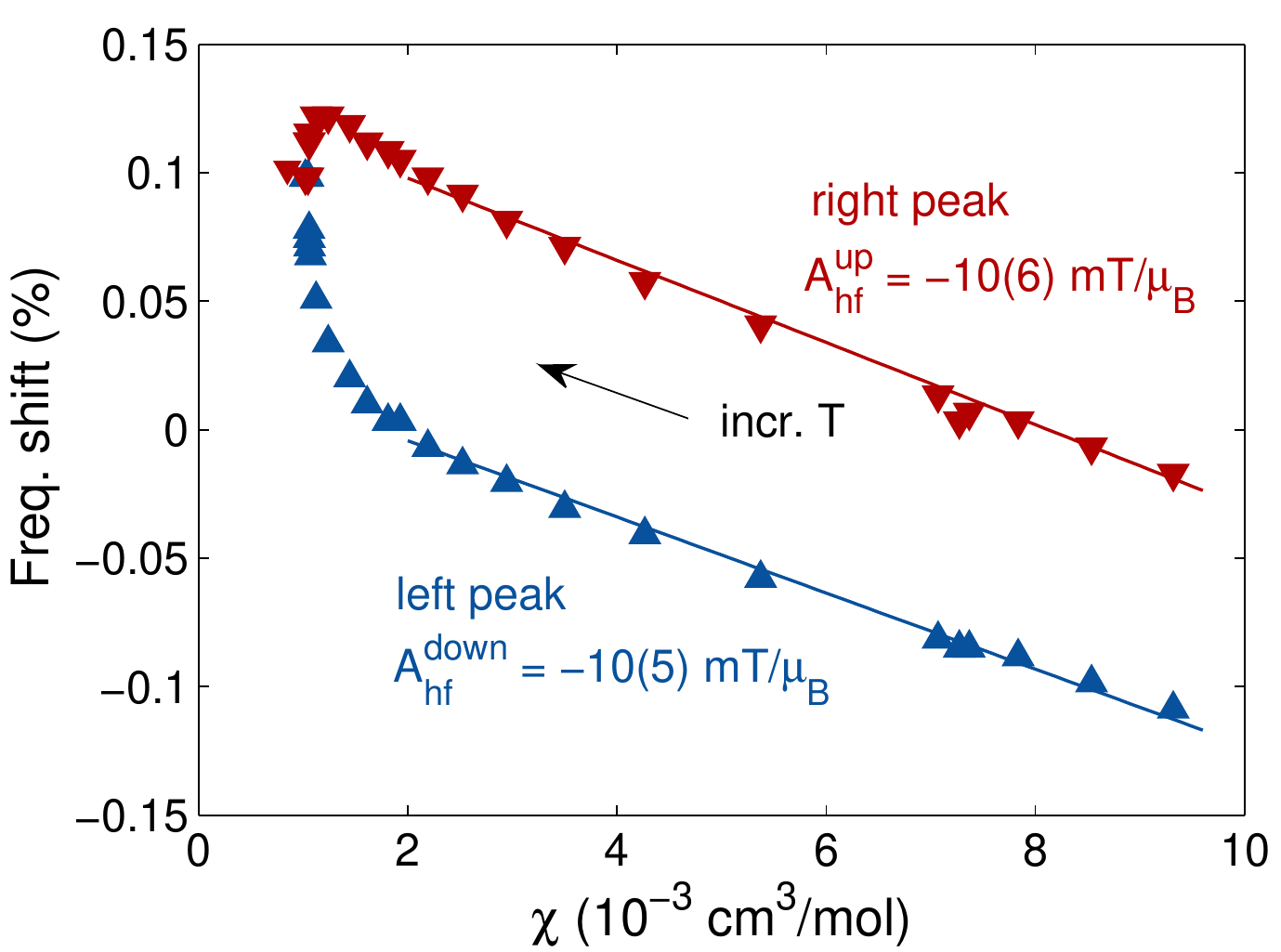}
  \caption{The Clogston-Jaccarino $K_s(T)$ vs.\ $\chi(T)$ plot for the 
  two inequivalent chlorine sites. The almost identical slopes result in 
  the same hyperfine coupling value, $A_\mathrm{hf} = -10(5)$\,mT/$\mu_{\mathrm{B}}$. 
The purely linear behavior is modified when the NMR signals start to merge. 
The arrow indicates the direction of increasing temperature.}
  \label{fig:CJ}
\end{figure}

By relating the frequency shifts $K_s(T)$ of the two $^{35}$Cl NMR lines 
to the susceptibility data $\chi(T)$ at the corresponding temperatures, 
we obtain the so-called Clogston-Jaccarino plot \cite{Clogston1962}, 
with $T$ as an implicit parameter.  
This is shown in Fig.~\ref{fig:CJ} and \tcr{was} 
used to evaluate the amplitude of the hyperfine interaction $A_{\mathrm{hf}}$. 
In our case, this is the same for both $^{35}$Cl sites, 
$A_{\mathrm{hf}} = -10(5)$\,mT/$\mu_\mathrm{B}$ (see parallel fit lines in 
Fig.~\ref{fig:CJ}), and was calculated 
by means of \cite{Shiroka2015} 
\begin{equation}
K_s = gA_{\mathrm{hf}}\chi + K_{\mathrm{orb}},
\label{eq:CJ}
\end{equation}
by assuming a standard electronic $g$ factor of 2.0. By extrapolating the straight line fits to $\chi_m=0$ 
(see Fig.~\ref{fig:CJ}), we obtain 
an orbital (i.e., temperature independent) shift $K_{\mathrm{orb}}$ of 
0.13\% for the upper line and of 0.03\% for the lower one. 
Above $T^{\star}$, the shift of the single NMR line extrapolates to 0.108\%, in 
excellent agreement with the
typical values of $^{35}$Cl chemical shift in 
the perchlorates (e.g., 0.105\% in KClO$_4$) \cite{Bryce2006}. 
While the latter reflects the undistorted geometry of the ClO$_4^{-}$ 
tetrahedra, the rather different $K_{\mathrm{orb}}$ values observed below 
$T^{\star}$, differing by approximately \ $+25\%$ and $-75\%$, respectively, 
are strong indicators of a significant deviation from the structural 
configuration adopted above $T^{\star}$. 
Considering that the $^{35}$Cl nuclei are located inside the tetrahedra, 
the change of $K_{\mathrm{orb}}$ across $T^{\star}$ has to be related to 
a static or dynamic modification of the tetrahedra, as confirmed by 
neutron diffraction experiments (see below).

\begin{figure}[htb]
\centering
  \includegraphics[width=0.45\textwidth]{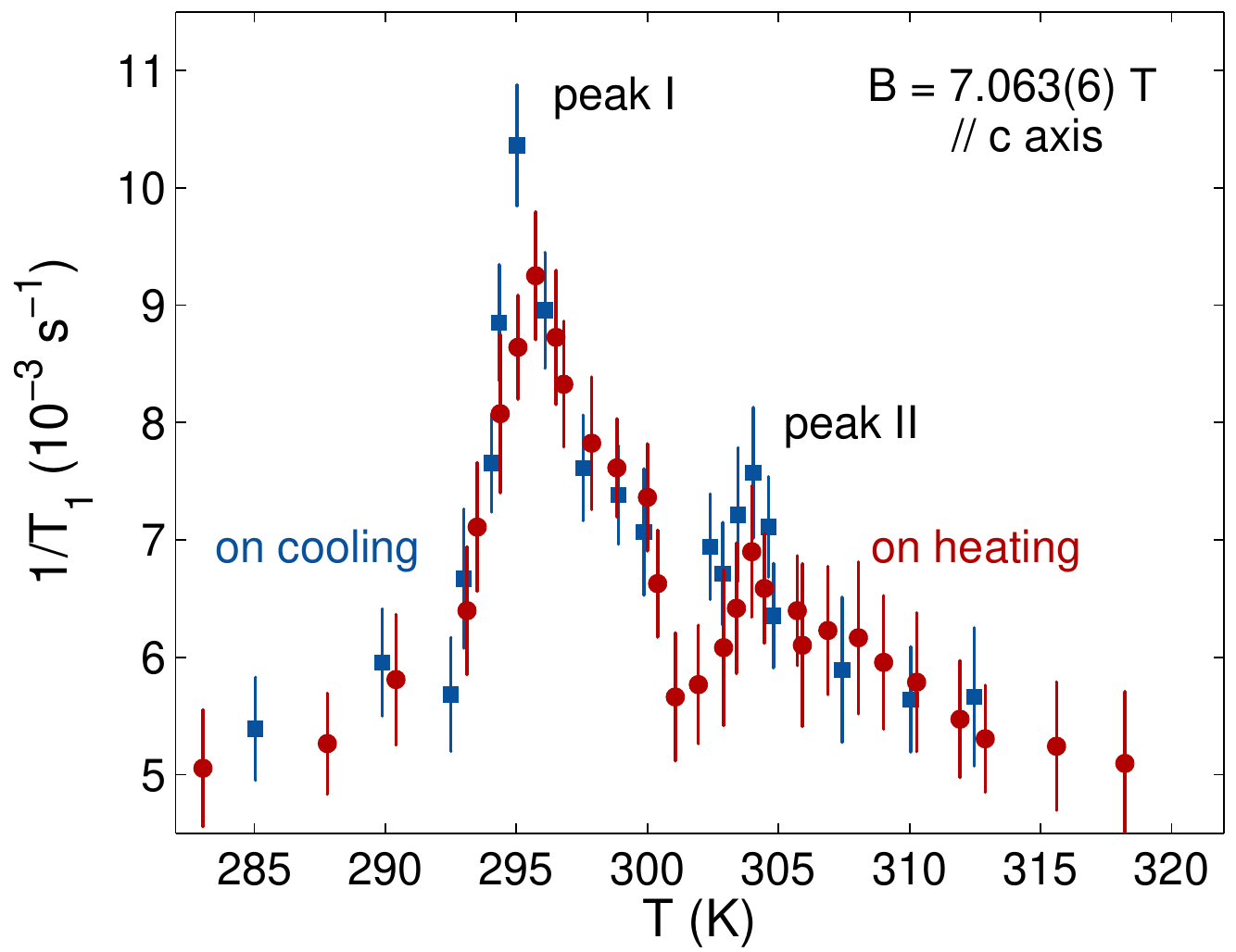}
  \caption{$^{35}$Cl NMR spin-lattice relaxation rate $1/T_1$ vs.\ 
  temperature measured upon heating (red circles) and cooling (blue 
  squares). Both datasets exhibit a main peak (I) at 295\,K 
  and a smaller secondary peak (II) at 304\,K. 
  To compensate for a known calibration offset, data measured upon 
  heating were shifted by $-2.0$\,K.}
  \label{fig:phase_trans}
\end{figure}

Since the NMR spin-lattice relaxation rates $1/T_1$ are known to be very 
sensitive to phase transitions \cite{Borsa2007}, \footnote{The 
sensitivity of the NMR relaxation rate to phase transitions arises 
from the so-called \emph{critical slowing down} of fluctuations in the proximity 
of a transition temperature $T_{c}$. The presence of long-range spatial 
correlations close to $T_{c}$ implies also slow time fluctuations. The 
latter, who match well the nuclear-spin time scales, provide a very 
effective relaxation channel and give rise to a peak in $1/T_{1}(T)$.} 
we performed detailed $T_{1}$ measurements in the temperature range 
between 280 and 320\,K. The relaxation-rate data collected by means of 
the inversion-recovery method (see Fig.~\ref{fig:phase_trans}), confirm the 
presence of an anomaly at 295(1)\,K, compatible with the $T^{\star}$ value,  
here defined as the merging point of the two distinct $^{35}$Cl NMR lines.
The $T_1$ values were determined from fits using the formula relevant for 
spin-3/2 nuclei \cite{Mcdowell1995}:
\begin{equation}
M_z(t) = M_0[1 - f\cdot(0.9\cdot e^{-(6t/T_1)^{\lambda}}+0.1 \cdot e^{-(t/T_1)^{\lambda}})].
\label{eq:T1_fit3_2}
\end{equation}
Here $M_0$ is the saturation value of nuclear magnetization, the parameter 
$f$ is ideally 2 for a full inversion, while $\lambda$ ($\leq 1$) is a stretching 
coefficient which accounts for a possible distribution of spin-lattice relaxation 
times around the mean $T_1$ value. \tcr{In our case, we find $f \sim 1.6(2)$ 
and $\lambda \sim 0.82(9)$ in the entire investigated temperature range. 
Both parameters exhibit a smooth variation across $T^*$, 
once more suggesting the structural character of the transition.} 
Since the two $^{35}$Cl NMR lines show very similar $T_{1}$ values, in the 
following we refer to only one of them, i.e., to the low-frequency signal.

The two datasets shown in Fig.~\ref{fig:phase_trans} correspond to the 
relaxation rates $1/T_{1}$ measured upon cooling (blue) and heating (red). 
In both cases a clear peak is observed at 295\,K, 
with a small yet distinct anomaly peaking at $T^{{\star}{\star}} = 304$\,K. 
The latter most likely reflects a minor change in the local atomic configuration, 
as suggested by neutron-diffraction data (see below). 
The lack of hysteresis, confirmed also by the previously mentioned heat-capacity data, 
is indicative of a second-order phase transition.
In the covered temperature range (up to 320\,K), no other peaks or anomalies 
were detected. Earlier NMR studies of pure pyrazine 
\cite{Schettino1972, Boyd1979} reported 
two anomalies  at similar temperatures (301.5\,K and 309\,K, respectively), 
\tcr{which were} interpreted as partial order-to-disorder transitions. 

\tcr{This similarity poses 
the question of whether this structural transition is also a feature of other 
molecular-copper complexes containing stacked pyrazine rings, 
such as the one-dimensional chains in Cu(pz)(NO$_3$)$_2$ \cite{Hammar1999} 
or the two-dimensional Cu(pz)$_2$X$_2$ family \cite{Choi2003} (here X indicates 
a counterion, e.g., ClO$_4^-$, BF$_4^-$, and PF$_6^-$).}
The role of counterions and ligands in \CUPZ\ was previously 
investigated by DFT calculations \cite{Vela2013}. 
These indicate that relevant structural changes with static or dynamic 
character may occur in \CUPZ, and consist of shearing-like distorsions 
of the pyrazine rings \cite{Vela2013}, in hydrogen-bonding 
effects \cite{Oneal2014}, or in the re-orientation, deformation, and 
freezing of the perchlorate counteranions \cite{Vela2013}. 
\tcr{The latter work reported a positive correlation between the 
intensity of AFM exchange interactions and the shearing-like distortions 
of pyrazine rings, suggesting that this may also be observed in other 
Cu(pz)-based magnets. Furthermore, if the presence of counterions 
is neglected, even tiny differences in the interatomic distances in the 
Cu-N$\cdots$N-Cu magnetic pathways give rise to sizable differences 
in the intensity of in-plane exchange interactions, 
regardless of the canting angle of the pyrazine ligands. 
Theoretical calculations considered also different types of counterions 
and concluded that the spin density on the central pyrazine and, hence, the 
magnetic exchange interaction, is enhanced by the more electronegative 
external atoms, as is the case of ClO$_4^-$.}
To assess whether, in our case, the detected phase transitions are due to 
one or more of the above mentioned changes predicted by DFT calculations, 
we had to perform neutron-diffraction experiments. 

\begin{figure}[htb]
\centering
  \includegraphics[width=0.48\textwidth]{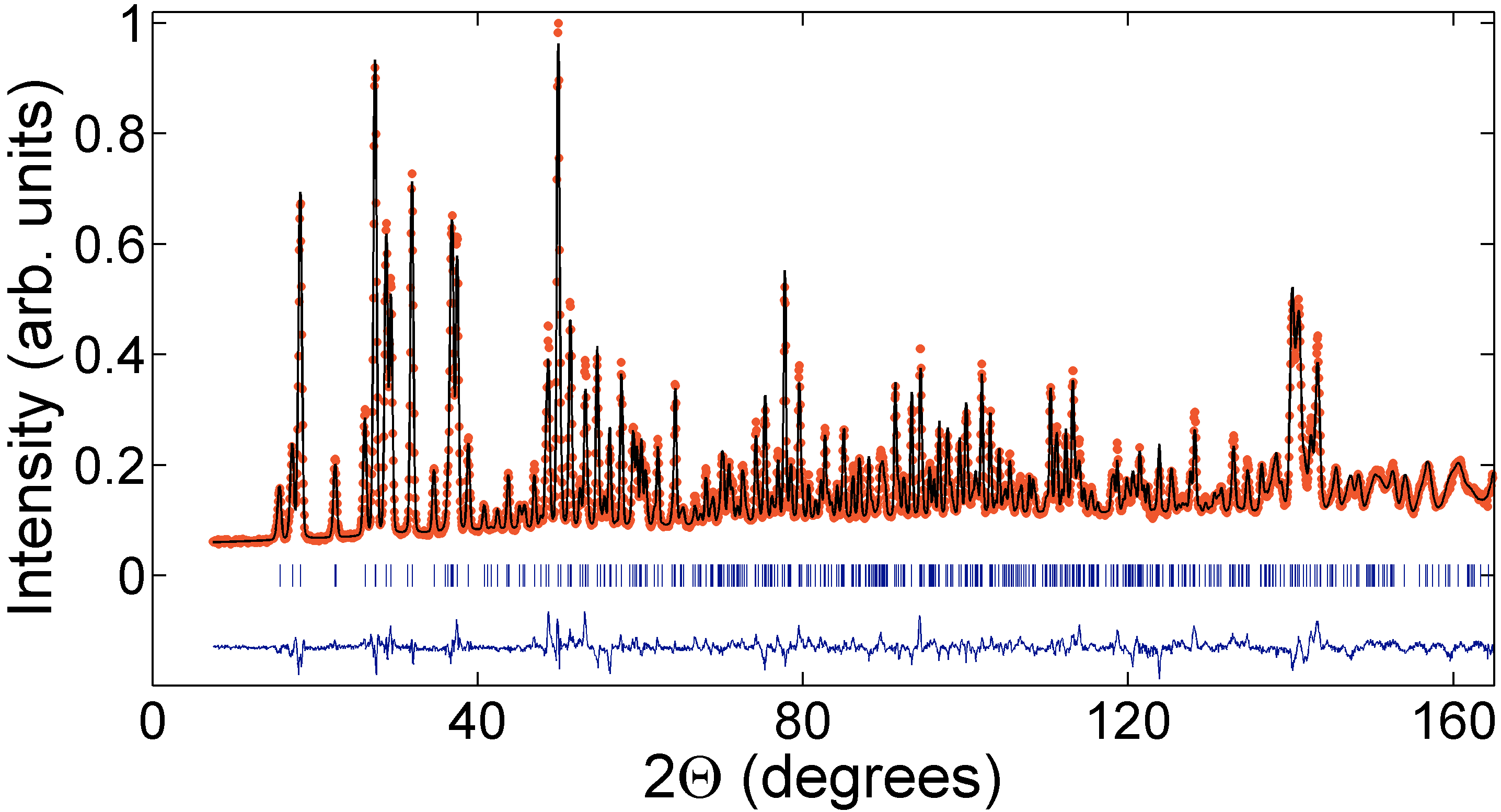}
  \caption{A typical neutron-diffraction pattern measured at 294\,K 
  using $\lambda$ = 1.8857\,\AA. The orange circles correspond to the 
  observed pattern and the black solid line to the Rietveld fit 
  obtained using the $C2/m$ space group. The positions of the Bragg 
  reflections and the difference between the observed and calculated 
  patterns are shown in blue.}
  \label{fig:rietveld}
\end{figure}
\begin{figure}[!htb]
\centering
\raisebox{-0.5\height}{\includegraphics[width=0.6\columnwidth,angle=0]{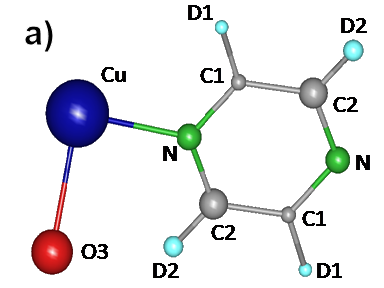}} 
\hfill 
\raisebox{-0.5\height}{\includegraphics[width=0.3\columnwidth,angle=0]{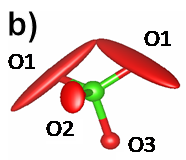}} 
\caption{\label{fig:struct_details}(a) Structural fragment of the 
pyrazine ring with labels identifying each atom. 
(b) Thermal ellipsoids of oxygen atoms in the ClO$^{-}_4$ perchlorate 
anions at 294\,K. Note the large displacements of the two O1 atoms.}
\end{figure}
%

\subsection{Neutron diffraction\label{ssec:neutron}}
The magnetization, specific-heat, and NMR results shown in the previous 
sections suggest a structural origin for the two transitions at 
$T^{\star} = 294$\,K and $T^{\star\star} = 304$\,K. To closely monitor 
the evolution of the crystal structure between 260 and 330\,K we 
performed systematic neutron powder-diffraction measurements. The initial 
structure refinement at 295\,K, i.e., just above the transition, was 
carried out by assuming as valid the $C2/m$ space group, previously 
proposed in the literature for the structure at 293\,K \cite{Darriet1979}. 
A representative Rietveld fit is shown in Fig.~\ref{fig:rietveld}. 

\begin{figure}[htb]
\centering
  \includegraphics[width=0.45\textwidth]{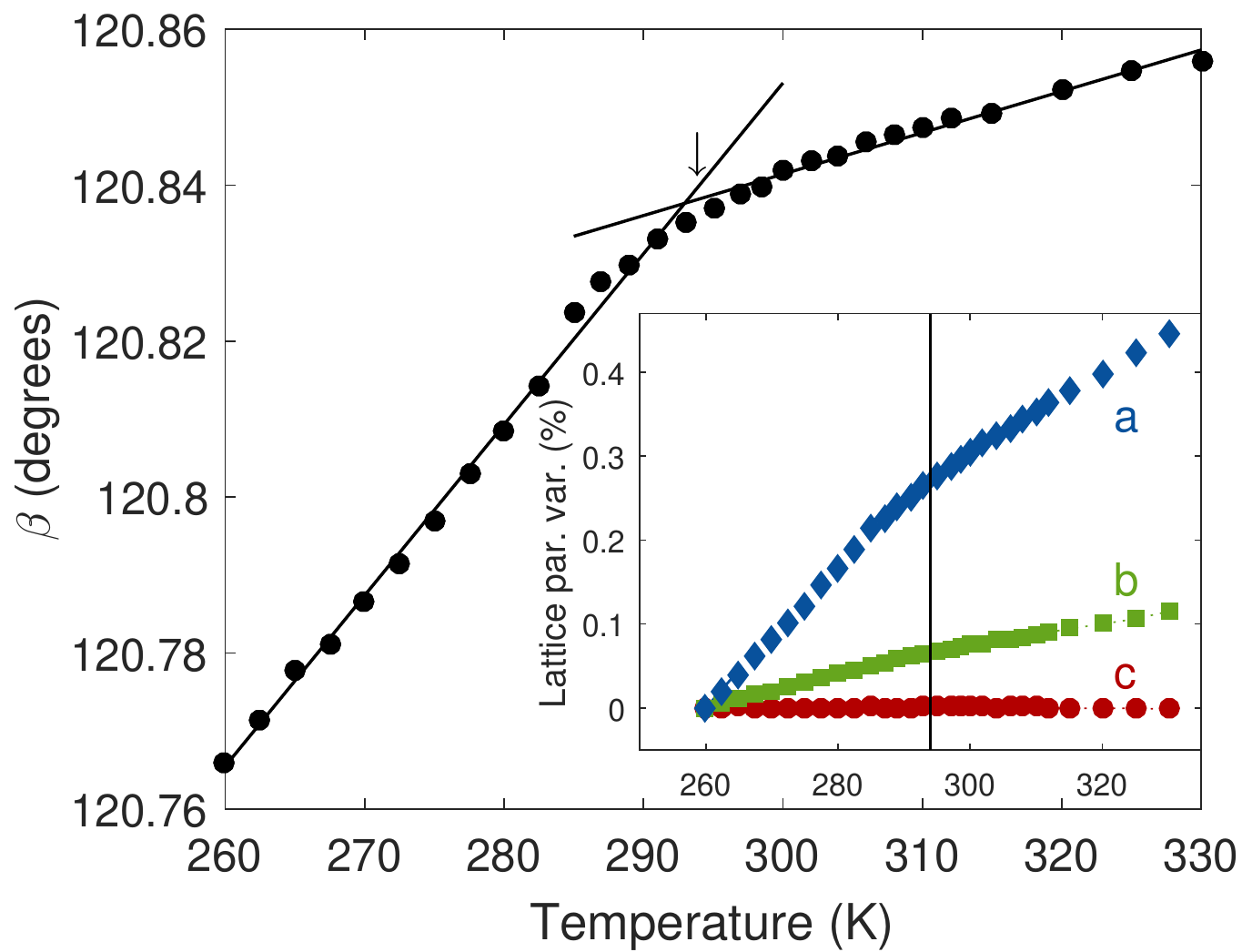}
  \caption{\label{fig:parameters}Relative variation of the lattice 
  parameters $a$, $b$, and $c$ (inset) and of the monoclinic angle 
  $\beta$ (main panel) with temperature. The interlayer axis ($c$) 
  is almost constant, the $b$ axis increases smoothly, whereas the 
  $a$ axis exhibits a tiny kink close to $T^{\star}$ (vertical line). 
  The monoclinic angle $\beta(T)$, too, shows a clear change of 
  slope across the transition (arrow).}
\end{figure}

The systematic extinctions \tcr{when using} this space group were found 
compatible with the high-resolution synchrotron x-ray powder diffraction 
data collected above and below the transition (260 and 295\,K), and 
refined by using the Le Bail method \cite{Lebail2005}. 
Besides $C2/m$, we also tried its subgroups $C2$ and $Cm$, 
indistinguishable in terms of solely systematic extinctions, but 
providing additional degrees of freedom, see Table~\ref{tab:structure}. 
The three groups were also tested with neutron data recorded at 260 
and 295\,K. Rietveld refinements using the $C2$ group, for which the 
absence of a mirror plane implies 4/4/2 different C/D/N Wyckoff positions 
(instead of 2/2/1 in the $C2/m$ case), provided slightly more realistic 
results for the environment of Cl in the perchlorate groups and better 
reliability indexes.  However, the larger number of free parameters 
resulted in the appearance of correlations and less stable fits. 
$C2/m$, which imposes additional restrictions, resulted in an improved 
fit stability and less correlations. Hence $C2/m$ was finally preferred 
for the description of the crystal structure across the full 260--330\,K 
temperature range. 
The obtained values of the atomic coordinates and Debye-Waller 
factors (see Figs.~\ref{fig:struct_details} and \ref{fig:ring_changes}), 
and in particular those of two of the oxygens in the perchlorate groups, 
should be considered with caution as they represent averages of the 
atomic positions in the actual (disordered) crystal structure. The resulting 
structural parameters and the agreement factors obtained at 295\,K 
with the two space groups are reported in Table~\ref{tab:structure} 
(see Appendix).

Figure~\ref{fig:parameters} shows the temperature evolution of the four 
lattice parameters ($a$, $b$, $c$, and the monoclinic angle $\beta$), 
as obtained from fits based on the $C2/m$ space group. 
While $\beta$ exhibits a clear change in slope close to 
$T^{\star}$, no evident features at the transition temperature were 
observed in the remaining lattice parameters, except for a tiny change 
in slope of $a(T)$ close to $T^{\star}$. From the NMR data, modifications 
related to the static (or dynamic) configuration of the perchlorate 
tetrahedra are expected.  
The above anomalies in the lattice parameters may also reflect 
the atomic rearrangements within the unit cell. 

In spite of the above mentioned restrictions, the structural parameters 
obtained from fits based on the $C2/m$ space group provide 
information not only on the atomic rearrangements, but also on the 
average structural disorder through the analysis of the Debye-Waller 
factors.  The latter quantify the mean-square displacements $U_{ij}$ 
(ADP, i.e., atomic displacement parameter) of each atom from its 
average position. In harmonic crystals, \tcr{$U_{ij}$}-s can be 
interpreted in terms of time-averaged mean-square displacements 
resulting from the normal modes of vibration. In disordered materials, 
however, they can also capture displacements due to positional disorder 
\cite{Medarde2013,Morin2016,Shang2018} \footnote{%
We recall that ADPs, which describe the anisotropic thermal motion 
through a symmetric rank-2 tensor, consist of six independent parameters. 
In the isotropic case, the off-diagonal terms clearly vanish. Most 
often $U_{ij}$ are used to express ADPs, since they represent directly 
the mean-square atomic displacements (some values relevant to our case 
are reported in Table~\ref{tab:structure}). An alternative, closely 
related quantity is $B_{ij} = 8\pi^2 U_{ij}$. Computer programs 
generally output the $\beta_{ij}$ values,  defined as 
$\beta_{ij} = B_{ij}x^{\star}_ix^{\star}_j/4$, since these minimize 
computation time and can be related directly to the reciprocal-cell 
parameters $x^{\star}_i$ and $x^{\star}_j$. The diagonal ADPs describe 
displacements along three mutually perpendicular axes of the ellipsoid 
and, hence, are always positive. On the other hand, since the other 
elements of the ADP tensor establish the orientation of the ellipsoid 
with respect to the crystal-lattice coordinate system, the off-diagonal 
elements can be either positive or negative, under the structural 
constrain $\beta_{ii}\beta_{jj} > \beta^2_{ij}$,  which is confirmed 
from our data and is a necessary requirement for the physical validity 
of the refinement.}. The $\beta_{ij}$ values, defined as 
$\beta_{ij} = 8\pi^2 U_{ij}x^{\star}_ix^{\star}_j/4$, with 
$x^{\star}_i$ and  $x^{\star}_j$ the reciprocal-cell parameters, were 
obtained from Rietveld fits with the FullProf suite package\cite{Carvajal1993}.
\begin{figure}[htb]
\centering
  \includegraphics[width=0.48\textwidth]{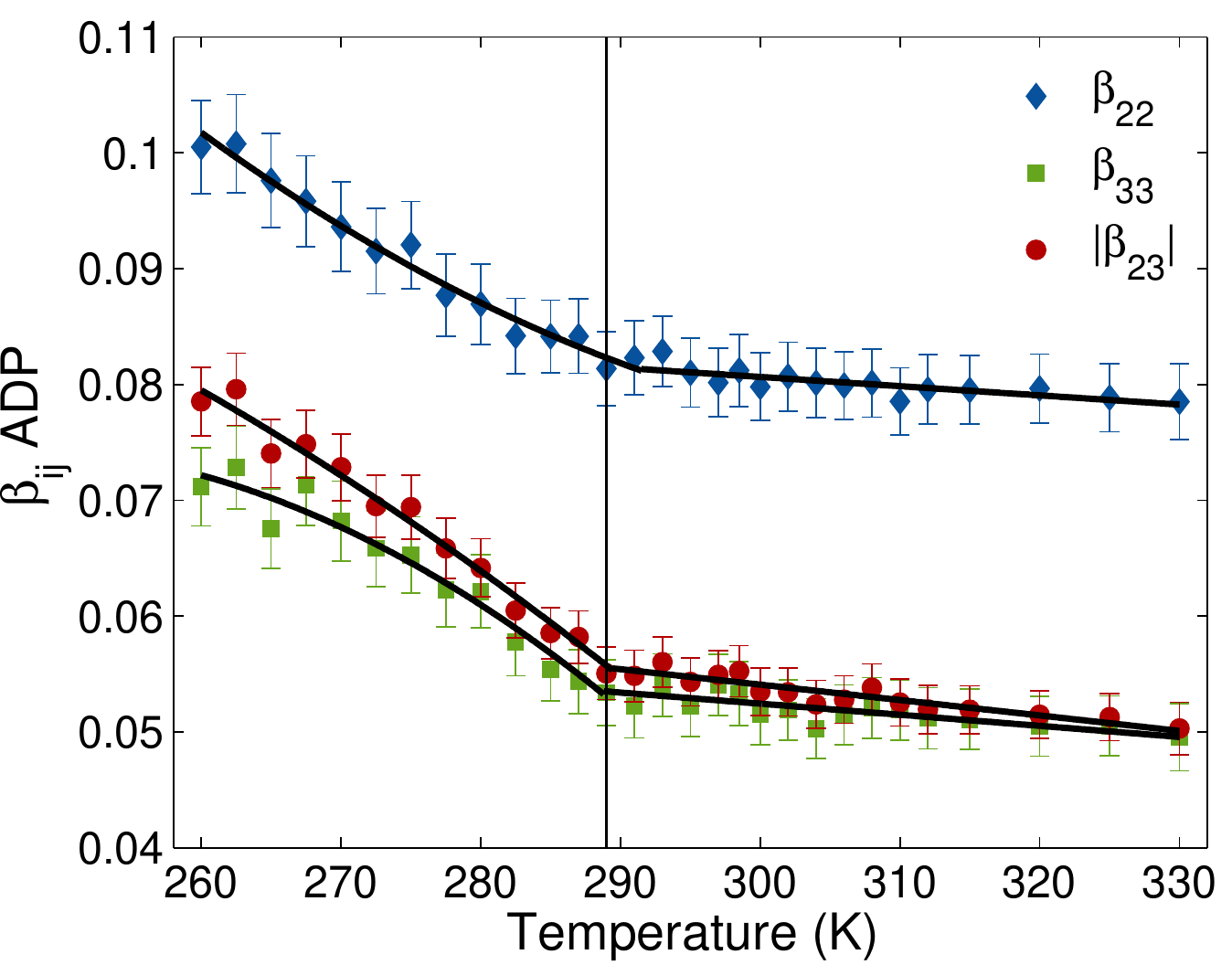}
  \caption{Temperature dependence of the diagonal $\beta_{22}$ (blue 
  diamonds) and $\beta_{33}$ (green squares) O1 ADPs and of the 
  absolute value of the off-diagonal $\beta_{23}$ term (red circles). 
  The crossing of the parabolic- and linear fits (black solid lines) 
  was used to better define the position of the anomaly at 
  $\sim 290$\,K (vertical line).}
  \label{fig:tetrahedron_changes}
\end{figure}

As shown in Fig.~\ref{fig:struct_details}(b), the thermal ellipsoids 
described by the ADPs are remarkably large for two of the three oxygen 
atoms that are not bound to a Cu magnetic site [here bound to O3 -- 
see Fig.~\ref{fig:struct_details}(a)]. 
The two oxygen atoms labeled O1 exhibit the largest 
displacements, and their $\beta_{22}$, $\beta_{33}$, and 
$\beta_{23}$ ADPs show clear anomalies across $T^{\star}$ 
(see Fig.~\ref{fig:tetrahedron_changes}). 
From 260\,K to 290\,K the diagonal terms $\beta_{22}$ and $\beta_{33}$ 
clearly decrease, to become almost constant above $T^{\star}$. The 
same reduction (in absolute value) is observed in the negative off-diagonal 
term $\beta_{23}$, which also saturates above 290\,K. Such general 
reductions of the absolute ADP values indicate a decrease of disorder 
above $T^{\star}$. 
The decrease of local disorder within the tetrahedra upon heating 
seems at first counterintuitive. Yet, it is in line with NMR results, 
which also indicate the number of inequivalent Cl sites (inside the 
ClO$^{4-}$ tetrahedra) to decrease upon heating from two to one 
at $T^{\star}$. A possible explanation for the enhanced degree of 
disorder at lower temperatures might be given by the coexistence of 
two competing oxygen positions. Note that significantly large ADP values 
persist also above $T^{\star}$, most likely indicating the presence of 
a residual positional disorder.

\begin{figure}[htb]
\centering
  \includegraphics[width=0.50\textwidth]{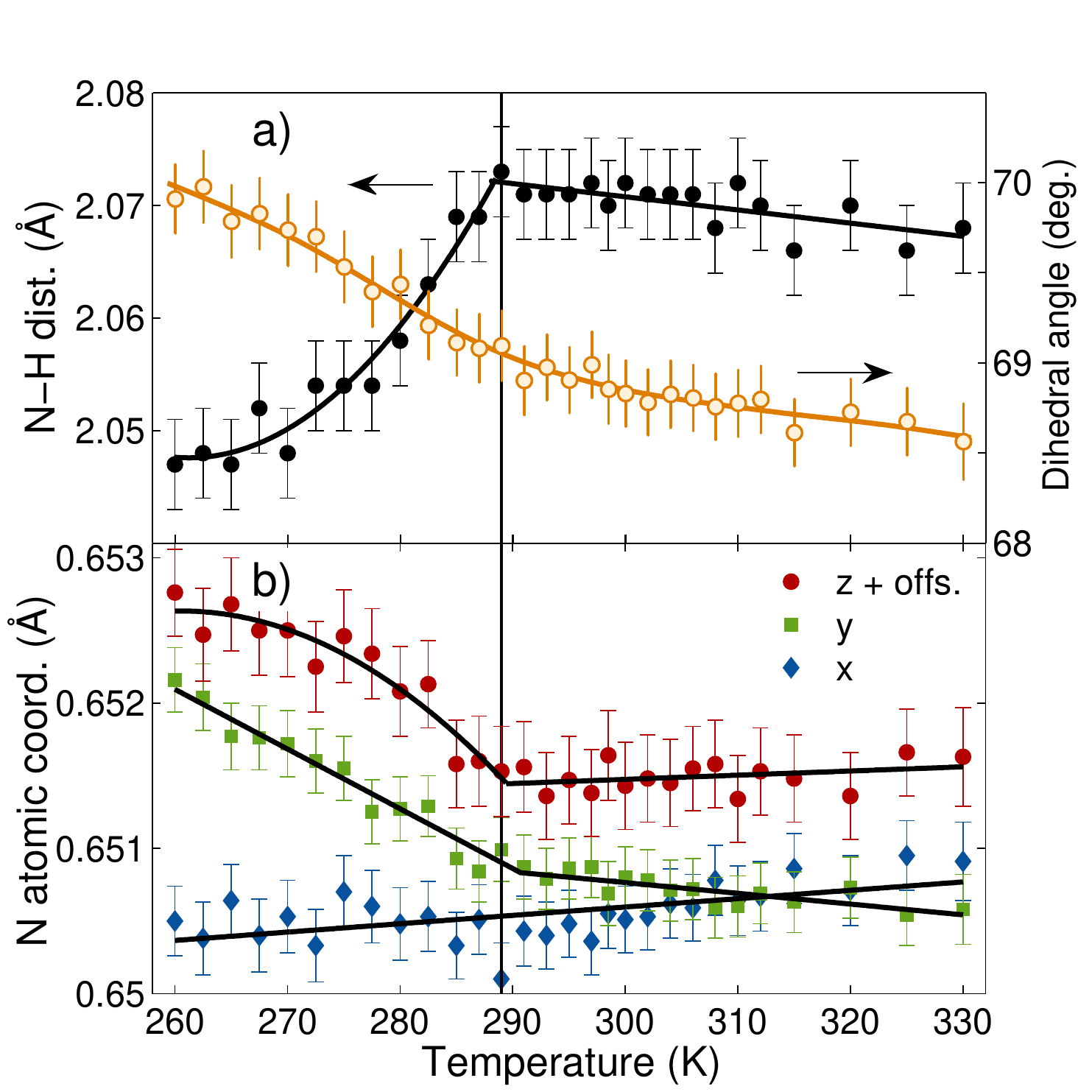}
  \caption{Temperature-dependence of the N--D1 distance (a) and of the 
  atomic coordinates of the N atoms (b). In all cases, the crossing of 
  parabolic- (linear-) and linear fits (black solid lines) 
  was used to define the position of the anomaly at $\sim 290$\,K 
  (vertical line). In (b), the $x$ and $y$ coordinates cross at about 
  308\,K. For clarity, the $z$ values were vertically offset by 0.647\,\AA.
  \tcr{The dihedral angle, also shown in (a), decreases with 
  temperature and exhibits an inflection point at $\sim 290$\,K.}}
  \label{fig:ring_changes}
\end{figure}
As shown in Fig.~\ref{fig:ring_changes}, the transition at $T^{\star}$ also 
involves subtle structural anomalies related to a distortion and 
reorientation of the four pyrazine rings. For instance, the 
N--D1 distance first increases up to 288\,K and then saturates, whereas the 
$y$ and $z$ atomic coordinates of the N atoms first rapidly decrease down 
to $T^{\star}$ and then continue decreasing with a reduced rate or saturate, 
respectively. We recall that $x$, $y$, and $z$ are the atomic coordinates, referred to the 
basis vectors $a$, $b$, and $c$, respectively.
As for the $x$ coordinate, it increases linearly across the covered 
temperature range, without showing any evident anomalies. Incidentally, 
the crossing of the $x$- and $y$ coordinate values at about 308\,K 
suggests a more regular in-plane arrangement of the pyrazine ligands 
above the second, yet much weaker, transition at $T^{\star\star}$. 
This corresponds to the pyrazine rings becoming closer to regular 
undeformed hexagons. \tcr{We note also that such rings are not 
perfectly flat. Nevertheless, considering the small deviatiations from an 
ideal plane, we still can define a dihedral angle 
$\theta_d$ between the average plane of the pyrazine 
rings and the basal plane of Cu octahedra. The temperature dependence 
of the dihedral angle 
is shown in Fig.~\ref{fig:ring_changes}(a) (right scale).  
This angle, very close to the values reported in the literature~\cite{Woodward2007}, 
decreases with temperature. Albeit weakly, the structural transition 
at $\sim 290$\,K is here reflected in an inflection point in $\theta_d(T)$.}

In any case, such modifications are subtle and 
modest in magnitude. We also note that the small mismatch between the 
transition temperature $T^{\star}$, established as outlined in 
Secs.~\ref{sec:res}B and \ref{sec:res}C, and the anomalies in the 
neutron-diffraction data is most likely due to temperature control and 
calibration issues, difficult to control exactly in experiments at room 
temperature. Moreover, we recall that the Cu--N and Cu--O3 distances 
do not exhibit clear anomalies at $T^{\star}$ and that the ratio 
$d$(Cu--N)/$d$(Cu--O3), reflecting the 
degree of the Jahn-Teller distortion of the octahedra centered at Cu 
sites, has a constant value of $1.14 \pm 0.08$. 
Finally, both the in-plane N--Cu--N- and the apical O--Cu--N angles are 
close to 90$^\circ$ and tend to the exact value of 90$^\circ$ upon 
increasing temperature. The absence of significant changes in these 
angles, which control the overlap of the CuN and CuO orbitals, indicates 
that the phase transition reported here does not imply sizable changes 
in the strength of the superexchange interactions.  

To provide a possible explanation for the above observations, we 
recall that structural phase transitions can be classified as either 
displacive or order-disorder, usually considered as mutually exclusive. 
The latter consists in the transition from a configuration where one 
or more sites exhibit split occupancies to a more ordered one \tcr{within} 
the  same set of atomic positions. From our neutron-diffraction and 
$C_p(T)$ measurements, we conclude that the observed transition is a 
second-order structural transition 
of the order-disorder type (with some possible residual displacive effects). 
In fact, below $T^{\star}$, two locally different Cl sites can be distinguished and, 
hence, the system is disordered. Above $T^{\star}$, despite the established 
equivalence of the two Cl sites, some residual disorder persists, most likely due to 
the co-existence of two similar configurations of the perchlorate subunits.

\vspace{-15pt}
\section{Conclusion\label{sec:Conclusion}}
By combining the results of magnetization-, specific-heat\mbox{-}, NMR-, 
and neutron-diffraction experiments, we identified a previously 
unnoticed second-order phase transition at $T^{\star}$ = 294\,K in 
\CUPZ\ and determined its order-to-disorder 
structural character. ${}^{35}$Cl NMR measurements indicate the 
presence of two inequivalent ${}^{35}$Cl sites which persists, 
upon heating, up to the transition temperature. Interestingly, the 
distance in frequency between the two ${}^{35}$Cl NMR lines starts 
reducing from 175\,K, i.e., the temperature at which a softening of 
the vibrational modes related to the pyrazine rings was previously observed. 
Above $T^{\star}$, the measured NMR orbital shift is compatible with 
typical values of the ${}^{35}$Cl nucleus in a tetrahedral environment. 
Below $T^{\star}$, instead, it assumes two significantly different 
values, with opposite signs for the two distinct sites, thus revealing 
strong deviations from the ideal tetrahedral geometry.  
The spin-lattice relaxation-rate data allowed us to precisely identify 
$T^{\star}$ and to observe also a minor anomaly at $T^{{\star}{\star}} = 304$\,K, 
attributable to secondary structural distortions. 
The refinement of the neutron-diffraction patterns, assuming the validity of the $C2/m$ 
space group, was used to monitor the temperature dependence of all the
structural parameters. A detailed analysis indicates that the reported 
transition affects the perchlorate ClO$^-_4$ tetrahedra, which adopt a 
more regular and ordered arrangement above $T^{\star}$. The relatively 
high values of the atomic displacements indicate the persistence of some 
positional disorder also above the transition, most likely due to 
the competition between two possible oxy\-gen\--si\-te positions, almost 
degenerate in energy. Similarly, reorentations of the pyrazine rings, 
occurring at the transition, are reflected in anomalies in the N--H 
distance and in the nitrogen atomic coordinates. As shown by 
magnetization- and specific-heat data, these subtle structural 
modifications do not affect the magnetic properties of \CUPZ\ 
in the covered temperature range. 
Future DFT calculations, which account for positional disorder 
and geometric distortions, are expected to provide further insight 
into the significance of the structural phase transition reported here.

\vspace{-15pt}
\section*{Acknowledgments}\vspace{-5pt}
This work is based on experiments performed at the Swiss spallation neutron source SINQ and the Swiss Light Source SLS, both at the Paul Scherrer Institute, Villigen, Switzerland. A special thank goes to N.\ Casati for supporting our preliminary x-ray 
experiments at the Materials Science Beamline (SLS). The authors thank M.\ Chinotti, D.\ Gawryluk, and M.\ Turnbull 
for helpful discussions and assistance. 
This work was financially supported in part by the Schweizerische 
Nationalfonds zur F\"{o}rderung der Wissenschaftlichen Forschung (SNF), 
Grant no.\ 200021-169455.

\section*{Appendix}
\subsection{Structural refinement comparison between the \texorpdfstring{$C2/m$}{C2/m} and \texorpdfstring{$C2$}{C2} 
space groups at \texorpdfstring{294\,K}{294 K}}

\begin{center}
\begin{longtable*}{lcc}
\hline\\[-2mm]   
\textrm{$T = 295$\,K} & \boldmath{$C2/m$} \textbf{model}   & \boldmath{$C2$} \textbf{model}  \ \\
\hline
\ \\[-2mm]
 $a$(\text{\AA})     & 9.78047(9)             & 9.78070(7)\ \\
 $b$(\text{\AA})     & 9.77594(9)             & 9.77589(7)\ \\
 $c$(\text{\AA})     & 8.16470(9)             & 8.16488(7)\ \\
 $\beta(^{\circ})$  & 120.8361(7)            & 120.8353(6)\ \\
\ \\[-2mm]
\hline\\[-2mm]
 \textbf{Cu} $(2a)/(2a)$                &\textrm{[0, 0, 0]}     &\textrm{[0, 0, 0]}\ \\[-0.2mm]
 $U_\mathrm{iso}(\text{\AA}^2)$    &  0.0118(9)                &  0.0126(9) \ \\
 \ \\
 \textbf{Cl} $(4i)/(4c)$                &\textrm{[0.2402(4), 0, 0.5391(5)]}     &\textrm{[0.2599(4),  0.4925(16),  0.4607(5)]}\ \\[-0.2mm]
 $U_\mathrm{iso}(\text{\AA}^2)$    &  0.012(3)               &  0.0292(10) \ \\
\ \\
 \textbf{O1} $(8j)/(4c)$                 &\textrm{[0.2052(9),  0.1075(9),  0.6031(11)]}     &\textrm{[0.3071(18),  0.588(3),  0.388(3)]}\ \\[-0.2mm]
 $U_{11}(\text{\AA}^2)$     & 0.057(5)              & 0.031(7) \ \\[-0.2mm]
 $U_{22}(\text{\AA}^2)$     & 0.393(14)            & 0.48(5) \ \\[-0.2mm]
 $U_{33}(\text{\AA}^2)$     & 0.128(6)             & 0.100(13) \ \\[-0.2mm]
 $U_{12}(\text{\AA}^2)$     & 0.061(6)             & $-$0.051(14) \ \\[-0.2mm]
 $U_{13}(\text{\AA}^2)$     & $-$0.030(4)            & $-$0.035(7) \ \\[-0.2mm]
 $U_{23}(\text{\AA}^2)$     & $-$0.195(7)            & 0.21(2) \ \\[-0.2mm]
 $U_\mathrm{iso}(\text{\AA}^2)$    &  0.193(8)                &  0.20(2) \ \\
  \ \\
  \textbf{O1'} $(-)/(4c)$                &\textrm{--}     &\textrm{[0.2713(19),  0.374(3),  0.397(3)]}\ \\[-0.2mm]
 $U_{11}(\text{\AA}^2)$     & --             & 0.061(10) \ \\[-0.2mm]
 $U_{22}(\text{\AA}^2)$     & --             & 0.217(16) \ \\[-0.2mm]
 $U_{33}(\text{\AA}^2)$     & --             & 0.176(18) \ \\[-0.2mm]
 $U_{12}(\text{\AA}^2)$     & --            & 0.006(10) \ \\[-0.2mm]
 $U_{13}(\text{\AA}^2)$     & --            & 0.012(11) \ \\[-0.2mm]
 $U_{23}(\text{\AA}^2)$     & --            & $-$0.173(14) \ \\[-0.2mm]
 $U_\mathrm{iso}(\text{\AA}^2)$    &  --               & 0.151(15) \ \\
  \ \\
  \textbf{O2} $(4i)/(4c)$                &\textrm{[0.4010(7),  0, 0.5868(9)]}     &\textrm{[0.1025(6),  0.504(2),  0.4146(9)]}\ \\[-0.2mm]
 $U_{11}(\text{\AA}^2)$     & --             & 0.015(3) \ \\[-0.2mm]
 $U_{22}(\text{\AA}^2)$     & --             & 0.069(5) \ \\[-0.2mm]
 $U_{33}(\text{\AA}^2)$     & --             & 0.067(5) \ \\[-0.2mm]
 $U_{12}(\text{\AA}^2)$     & --            & 0.005(6) \ \\[-0.2mm]
  $U_{13}(\text{\AA}^2)$     & --            & 0.009(3) \ \\[-0.2mm]
  $U_{23}(\text{\AA}^2)$     & --            & 0.003(7) \ \\[-0.2mm]
 $U_\mathrm{iso}(\text{\AA}^2)$    &  0.042(4)               &  0.047(4) \ \\
 \ \\
  \textbf{O3} $(4i)/(4c)$                &\textrm{[0.1363(7),  0,  0.3366(8)]}     &\textrm{[0.3643(6),  0.503(2),  0.6617(8)]}\ \\[-0.2mm]
 $U_\mathrm{iso}(\text{\AA}^2)$    &  0.0311(13)                &  0.0335(13) \ \\
 \ \\
  \textbf{C1} $(8j)/(4c)$                &\textrm{[0.8968(3),  0.7691(3),  0.1448(4)]}     &\textrm{[0.8948(10)  0.2307(14)  0.1506(11)]}\ \\[-0.2mm]
 $U_\mathrm{iso}(\text{\AA}^2)$    &  0.0145(7)               &  0.0168(5)$^*$ \ \\
 \ \\
  \textbf{C2} $(8j)/(4c)$                &\textrm{[0.7985(3),  0.6701(3),  0.1524(4)]}     &\textrm{[0.2996(9)  0.8400(15)  0.1504(12)]}\ \\[-0.2mm]
 $U_\mathrm{iso}(\text{\AA}^2)$    & 0.0190(7)               &  0.0168(5)$^*$  \ \\
 \ \\
  \textbf{C3} $(-)/ (4c)$                &\textrm{--}     &\textrm{[0.0997(10),  0.7683(14),  0.8572(11)]}\ \\[-0.2mm]
 $U_\mathrm{iso}(\text{\AA}^2)$    &  --               & 0.0168(5)$^*$ \ \\
 \ \\
  \textbf{C4} $(-)/ (4c)$                &\textrm{--}     &\textrm{[0.7037(9),  0.1779(14),  0.8445(12)]}\ \\[-0.2mm]
 $U_\mathrm{iso}(\text{\AA}^2)$    & --               &  0.0168(5)$^*$ \ \\
 \ \\
  \textbf{D1} $(8j)/(4c)$                &\textrm{[1.0197(5),  0.7807(4),  0.2700(6)]}     &\textrm{[0.0283(9),  0.2351(14),  0.2558(9)]}\ \\[-0.2mm]
 \textit{Occ}                 &    0.978(2)$^*$                    &  0.961(7)$^*$   \ \\[-0.2mm]
 $U_\mathrm{iso}(\text{\AA}^2)$    & 0.0428(12)                &  0.0254(9)$^*$ \ \\
 \ \\
  \textbf{D2} $(8j)/(4c)$                &\textrm{[0.8379(5),  0.6088(4),  0.2768(5)]}     &\textrm{[0.3453(8),  0.9067(14),  0.2672(9)]}\ \\[-0.2mm]
 \textit{Occ}                 &    0.978(2)$^*$                     &  0.961(7)$^*$   \ \\[-0.2mm]
 $U_\mathrm{iso}(\text{\AA}^2)$    &  0.0372(10)               &  0.0254(9)$^*$ \ \\
 \ \\
  \textbf{D3} $(-)/(4c)$                &\textrm{--}     &\textrm{[0.9892(8),  0.7973(14),  0.7166(10)]}\ \\[-0.2mm]
 \textit{Occ}                 &    --                   &  0.961(7)$^*$   \ \\[-0.2mm]
 $U_\mathrm{iso}(\text{\AA}^2)$    & --                &  0.0254(9)$^*$ \ \\
 \ \\
  \textbf{D4} $(-)/(4c)$                &\textrm{--}     &\textrm{[0.6732(8),  0.1254(14),  0.7153(10)]}\ \\[-0.2mm]
 \textit{Occ}                 &    --                   &  0.961(7)$^*$   \ \\[-0.2mm]
 $U_\mathrm{iso}(\text{\AA}^2)$    &  --                & 0.0254(9)$^*$ \ \\
 \ \\
   \textbf{N1} $(8j)/(4c)$                &\textrm{[0.6505(2),  0.65082(20),  0.0041(3)]}     &\textrm{[0.8503(6),  0.1551(15),  0.9987(8)]}\ \\[-0.2mm]
 $U_\mathrm{iso}(\text{\AA}^2)$    &  0.0096(4)                &  0.0141(5)$^*$ \ \\
 \ \\
   \textbf{N2} $(-)/(4c)$                &\textrm{--}     &\textrm{[0.1512(6),  0.8534(15),  0.0060(8)]}\ \\[-0.2mm]
 $U_\mathrm{iso}(\text{\AA}^2)$    &  --               & 0.0141(5)$^*$ \ \\
\ \\[-2mm]
\hline\\[-2mm]

 \textbf{$\lambda$(\,\AA)   }                    & \textrm{\textbf{1.494 1.886}}         & \textrm{\textbf{1.494 1.886}}\ \\
 \textbf{$\chi^2$}                                    & 5.18 8.54                                       & 3.27 5.06\ \\
 $R_{exp}$                                              & 4.50 3.49                                       & 4.48 3.48\ \\
 $R_{wp}$                                               & 10.2 10.2                                      & 8.10 7.82\ \\
 $R_p$                                                    & 9.37 9.14                                      & 7.53 7.25\ \\
 \hline
\ \\
\caption[width=0.5]{\label{tab:structure} Structural parameters as determined 
from neutron powder diffraction at $T= 295$\,K, using the space 
groups $C2/m$ and $C2$, respectively \footnotemark[3].
For the last group, that is non-centrosymmetric, the origin is floating. 
For that reason the $y$-coordinate of Cu has been fixed to zero, 
matching the Cu postion in $C2/m$. To further facilitate comparisons, 
the setting for the monoclinic cell was chosen to be the same as in 
reference \cite{Woodward2007}. For every space group the fits were 
carried out using a single structural model to refine simultaneosly 
the data recorded at $\lambda = 1.494$  and 1.886 \text{\AA}. The 
related .cif file with the refinement parameters can be found in the 
Supplemental Material \footnotemark[4].}
\footnotetext[3]{The quantities marked with (*) in Table I were 
constrained to be identical for the same atomic species. The occupation 
of the D sites was found to be very close to 100\%. The occupation of 
the remaining atomic sites was 100\%}. 
\footnotetext[4]{The Supplemental Material available at [URL] includes the two 
structural .cif files used for refining the neutron-diffraction patterns at 295 K, 
by assuming as valid the space groups $C2$ and $C2/m$, respectively.}
\end{longtable*}
\end{center}


\begin{thebibliography}{29}%
\makeatletter
\providecommand \@ifxundefined [1]{%
 \@ifx{#1\undefined}
}%
\providecommand \@ifnum [1]{%
 \ifnum #1\expandafter \@firstoftwo
 \else \expandafter \@secondoftwo
 \fi
}%
\providecommand \@ifx [1]{%
 \ifx #1\expandafter \@firstoftwo
 \else \expandafter \@secondoftwo
 \fi
}%
\providecommand \natexlab [1]{#1}%
\providecommand \enquote  [1]{``#1''}%
\providecommand \bibnamefont  [1]{#1}%
\providecommand \bibfnamefont [1]{#1}%
\providecommand \citenamefont [1]{#1}%
\providecommand \href@noop [0]{\@secondoftwo}%
\providecommand \href [0]{\begingroup \@sanitize@url \@href}%
\providecommand \@href[1]{\@@startlink{#1}\@@href}%
\providecommand \@@href[1]{\endgroup#1\@@endlink}%
\providecommand \@sanitize@url [0]{\catcode `\\12\catcode `\$12\catcode
  `\&12\catcode `\#12\catcode `\^12\catcode `\_12\catcode `\%12\relax}%
\providecommand \@@startlink[1]{}%
\providecommand \@@endlink[0]{}%
\providecommand \url  [0]{\begingroup\@sanitize@url \@url }%
\providecommand \@url [1]{\endgroup\@href {#1}{\urlprefix }}%
\providecommand \urlprefix  [0]{URL }%
\providecommand \Eprint [0]{\href }%
\providecommand \doibase [0]{https://doi.org/}%
\providecommand \selectlanguage [0]{\@gobble}%
\providecommand \bibinfo  [0]{\@secondoftwo}%
\providecommand \bibfield  [0]{\@secondoftwo}%
\providecommand \translation [1]{[#1]}%
\providecommand \BibitemOpen [0]{}%
\providecommand \bibitemStop [0]{}%
\providecommand \bibitemNoStop [0]{.\EOS\space}%
\providecommand \EOS [0]{\spacefactor3000\relax}%
\providecommand \BibitemShut  [1]{\csname bibitem#1\endcsname}%
\let\auto@bib@innerbib\@empty
\bibitem [{\citenamefont {Darriet}\ \emph {et~al.}(1979)\citenamefont
  {Darriet}, \citenamefont {Haddad}, \citenamefont {Duesler},\ and\
  \citenamefont {Hendrickson}}]{Darriet1979}%
  \BibitemOpen
  \bibfield  {author} {\bibinfo {author} {\bibfnamefont {J.}~\bibnamefont
  {Darriet}}, \bibinfo {author} {\bibfnamefont {M.~S.}\ \bibnamefont {Haddad}},
  \bibinfo {author} {\bibfnamefont {E.~N.}\ \bibnamefont {Duesler}},\ and\
  \bibinfo {author} {\bibfnamefont {D.~N.}\ \bibnamefont {Hendrickson}},\
  }\bibfield  {title} {\bibinfo {title} {Crystal structure and magnetic
  properties of bis\-(py\-ra\-zi\-ne)\-cop\-per({II}) per\-chlorate,
  {C}u(pyz)$_2$({C}l{O}$_4$)$_2$, a two-dimensional {H}eisenberg
  antiferromagnet},\ }\href {https://doi.org/10.1021/ic50200a008} {\bibfield
  {journal} {\bibinfo  {journal} {Inorg. Chem.}\ }\textbf {\bibinfo {volume}
  {18}},\ \bibinfo {pages} {2679} (\bibinfo {year} {1979})}\BibitemShut
  {NoStop}%
\bibitem [{\citenamefont {Choi}\ \emph {et~al.}(2003)\citenamefont {Choi},
  \citenamefont {Woodward}, \citenamefont {Musfeldt}, \citenamefont {Landee},\
  and\ \citenamefont {Turnbull}}]{Choi2003}%
  \BibitemOpen
  \bibfield  {author} {\bibinfo {author} {\bibfnamefont {J.}~\bibnamefont
  {Choi}}, \bibinfo {author} {\bibfnamefont {J.~D.}\ \bibnamefont {Woodward}},
  \bibinfo {author} {\bibfnamefont {J.~L.}\ \bibnamefont {Musfeldt}}, \bibinfo
  {author} {\bibfnamefont {C.~P.}\ \bibnamefont {Landee}},\ and\ \bibinfo
  {author} {\bibfnamefont {M.~M.}\ \bibnamefont {Turnbull}},\ }\bibfield
  {title} {\bibinfo {title} {Vibrational properties of
  {C}u(pz)$_2$({C}l{O}$_4$)$_2$: Evidence for enhanced low-temperature hydrogen
  bonding in square ${S} = 1/2$ molecular antiferromagnets},\ }\href
  {https://doi.org/10.1021/cm030049e} {\bibfield  {journal} {\bibinfo
  {journal} {Chem. Mater.}\ }\textbf {\bibinfo {volume} {15}},\ \bibinfo
  {pages} {2797} (\bibinfo {year} {2003})}\BibitemShut {NoStop}%
\bibitem [{\citenamefont {Landee}\ and\ \citenamefont
  {M.Turnbull}(2013)}]{Landee2013}%
  \BibitemOpen
  \bibfield  {author} {\bibinfo {author} {\bibfnamefont {C.~P.}\ \bibnamefont
  {Landee}}\ and\ \bibinfo {author} {\bibfnamefont {M.}~\bibnamefont
  {M.Turnbull}},\ }\bibfield  {title} {\bibinfo {title} {Recent developments in
  low-dimensional copper({II}) molecular magnets},\ }\href
  {https://doi.org/10.1002/ejic.201300133} {\bibfield  {journal} {\bibinfo
  {journal} {Eur. J. Inorg. Chem.}\ }\textbf {\bibinfo {volume} {13}},\
  \bibinfo {pages} {2266} (\bibinfo {year} {2013})}\BibitemShut {NoStop}%
\bibitem [{\citenamefont {Tsyrulin}\ \emph {et~al.}(2010)\citenamefont
  {Tsyrulin}, \citenamefont {Xiao}, \citenamefont {Schneidewind}, \citenamefont
  {Link}, \citenamefont {R\o{}nnow}, \citenamefont {Gavilano}, \citenamefont
  {Landee}, \citenamefont {Turnbull},\ and\ \citenamefont
  {Kenzelmann}}]{Tsyrulin2010}%
  \BibitemOpen
  \bibfield  {author} {\bibinfo {author} {\bibfnamefont {N.}~\bibnamefont
  {Tsyrulin}}, \bibinfo {author} {\bibfnamefont {F.}~\bibnamefont {Xiao}},
  \bibinfo {author} {\bibfnamefont {A.}~\bibnamefont {Schneidewind}}, \bibinfo
  {author} {\bibfnamefont {P.}~\bibnamefont {Link}}, \bibinfo {author}
  {\bibfnamefont {H.~M.}\ \bibnamefont {R\o{}nnow}}, \bibinfo {author}
  {\bibfnamefont {J.}~\bibnamefont {Gavilano}}, \bibinfo {author}
  {\bibfnamefont {C.~P.}\ \bibnamefont {Landee}}, \bibinfo {author}
  {\bibfnamefont {M.~M.}\ \bibnamefont {Turnbull}},\ and\ \bibinfo {author}
  {\bibfnamefont {M.}~\bibnamefont {Kenzelmann}},\ }\bibfield  {title}
  {\bibinfo {title} {Two-dimensional square-lattice ${S}=1/2$ antiferromagnet
  $\text{Cu}{(\text{pz})}_{2}{({\text{ClO}}_{4})}_{2}$},\ }\href
  {https://doi.org/10.1103/PhysRevB.81.134409} {\bibfield  {journal} {\bibinfo
  {journal} {Phys. Rev. B}\ }\textbf {\bibinfo {volume} {81}},\ \bibinfo
  {pages} {134409} (\bibinfo {year} {2010})}\BibitemShut {NoStop}%
\bibitem [{\citenamefont {Lancaster}\ \emph {et~al.}(2007)\citenamefont
  {Lancaster}, \citenamefont {Blundell}, \citenamefont {Brooks}, \citenamefont
  {Baker}, \citenamefont {Pratt}, \citenamefont {Manson}, \citenamefont
  {Conner}, \citenamefont {Xiao}, \citenamefont {Landee}, \citenamefont
  {Chaves}, \citenamefont {Soriano}, \citenamefont {Novak}, \citenamefont
  {Papageorgiou}, \citenamefont {Bianchi}, \citenamefont {Herrmannsd\"orfer},
  \citenamefont {Wosnitza},\ and\ \citenamefont {Schlueter}}]{Lancaster2007}%
  \BibitemOpen
  \bibfield  {author} {\bibinfo {author} {\bibfnamefont {T.}~\bibnamefont
  {Lancaster}}, \bibinfo {author} {\bibfnamefont {S.~J.}\ \bibnamefont
  {Blundell}}, \bibinfo {author} {\bibfnamefont {M.~L.}\ \bibnamefont
  {Brooks}}, \bibinfo {author} {\bibfnamefont {P.~J.}\ \bibnamefont {Baker}},
  \bibinfo {author} {\bibfnamefont {F.~L.}\ \bibnamefont {Pratt}}, \bibinfo
  {author} {\bibfnamefont {J.~L.}\ \bibnamefont {Manson}}, \bibinfo {author}
  {\bibfnamefont {M.~M.}\ \bibnamefont {Conner}}, \bibinfo {author}
  {\bibfnamefont {F.}~\bibnamefont {Xiao}}, \bibinfo {author} {\bibfnamefont
  {C.~P.}\ \bibnamefont {Landee}}, \bibinfo {author} {\bibfnamefont {F.~A.}\
  \bibnamefont {Chaves}}, \bibinfo {author} {\bibfnamefont {S.}~\bibnamefont
  {Soriano}}, \bibinfo {author} {\bibfnamefont {M.~A.}\ \bibnamefont {Novak}},
  \bibinfo {author} {\bibfnamefont {T.~P.}\ \bibnamefont {Papageorgiou}},
  \bibinfo {author} {\bibfnamefont {A.~D.}\ \bibnamefont {Bianchi}}, \bibinfo
  {author} {\bibfnamefont {T.}~\bibnamefont {Herrmannsd\"orfer}}, \bibinfo
  {author} {\bibfnamefont {J.}~\bibnamefont {Wosnitza}},\ and\ \bibinfo
  {author} {\bibfnamefont {J.~A.}\ \bibnamefont {Schlueter}},\ }\bibfield
  {title} {\bibinfo {title} {Magnetic order in the ${S} = 1/2$ two-dimensional
  molecular antiferromagnet copper pyrazine perchlorate
  {C}u(pz)$_2$({Cl}{O}$_4$)$_2$},\ }\href
  {https://doi.org/10.1103/PhysRevB.75.094421} {\bibfield  {journal} {\bibinfo
  {journal} {Phys. Rev. B}\ }\textbf {\bibinfo {volume} {75}},\ \bibinfo
  {pages} {094421} (\bibinfo {year} {2007})}\BibitemShut {NoStop}%
\bibitem [{\citenamefont {Barbero}\ \emph {et~al.}(2016)\citenamefont
  {Barbero}, \citenamefont {Shiroka}, \citenamefont {Landee}, \citenamefont
  {Pikulski}, \citenamefont {Ott},\ and\ \citenamefont {Mesot}}]{Barbero2016}%
  \BibitemOpen
  \bibfield  {author} {\bibinfo {author} {\bibfnamefont {N.}~\bibnamefont
  {Barbero}}, \bibinfo {author} {\bibfnamefont {T.}~\bibnamefont {Shiroka}},
  \bibinfo {author} {\bibfnamefont {C.~P.}\ \bibnamefont {Landee}}, \bibinfo
  {author} {\bibfnamefont {M.}~\bibnamefont {Pikulski}}, \bibinfo {author}
  {\bibfnamefont {H.-R.}\ \bibnamefont {Ott}},\ and\ \bibinfo {author}
  {\bibfnamefont {J.}~\bibnamefont {Mesot}},\ }\bibfield  {title} {\bibinfo
  {title} {Pressure and magnetic field effects on a quasi-two-dimensional
  spin-$1/2$ {H}eisenberg antiferromagnet},\ }\href
  {https://doi.org/10.1103/PhysRevB.93.054425} {\bibfield  {journal} {\bibinfo
  {journal} {Phys. Rev. B}\ }\textbf {\bibinfo {volume} {93}},\ \bibinfo
  {pages} {054425} (\bibinfo {year} {2016})}\BibitemShut {NoStop}%
\bibitem [{\citenamefont {Mermin}\ and\ \citenamefont
  {Wagner}(1966)}]{Mermin1966}%
  \BibitemOpen
  \bibfield  {author} {\bibinfo {author} {\bibfnamefont {N.~D.}\ \bibnamefont
  {Mermin}}\ and\ \bibinfo {author} {\bibfnamefont {H.}~\bibnamefont
  {Wagner}},\ }\bibfield  {title} {\bibinfo {title} {Absence of ferromagnetism
  or antiferromagnetism in one- or two-dimensional isotropic {H}eisenberg
  models},\ }\href {https://doi.org/10.1103/PhysRevLett.17.1133} {\bibfield
  {journal} {\bibinfo  {journal} {Phys. Rev. Lett.}\ }\textbf {\bibinfo
  {volume} {17}},\ \bibinfo {pages} {1133} (\bibinfo {year}
  {1966})}\BibitemShut {NoStop}%
\bibitem [{\citenamefont {Vela}\ \emph {et~al.}(2013)\citenamefont {Vela},
  \citenamefont {Jornet-Somoza}, \citenamefont {Turnbull}, \citenamefont
  {Feyerherm}, \citenamefont {Novoa},\ and\ \citenamefont {Deumal}}]{Vela2013}%
  \BibitemOpen
  \bibfield  {author} {\bibinfo {author} {\bibfnamefont {S.}~\bibnamefont
  {Vela}}, \bibinfo {author} {\bibfnamefont {J.}~\bibnamefont {Jornet-Somoza}},
  \bibinfo {author} {\bibfnamefont {M.~M.}\ \bibnamefont {Turnbull}}, \bibinfo
  {author} {\bibfnamefont {R.}~\bibnamefont {Feyerherm}}, \bibinfo {author}
  {\bibfnamefont {J.~J.}\ \bibnamefont {Novoa}},\ and\ \bibinfo {author}
  {\bibfnamefont {M.}~\bibnamefont {Deumal}},\ }\bibfield  {title} {\bibinfo
  {title} {Dividing the spoils: Role of pyrazine ligands and perchlorate
  counterions in the magnetic properties of
  bis\-(py\-ra\-zine)\-di\-per\-chlorate copper({II}),
  [{C}u(pz)$_2$]({C}l{O}$_4$)$_2$},\ }\href {https://doi.org/10.1021/ic400712s}
  {\bibfield  {journal} {\bibinfo  {journal} {Inorg. Chem.}\ }\textbf {\bibinfo
  {volume} {52}},\ \bibinfo {pages} {12923} (\bibinfo {year}
  {2013})}\BibitemShut {NoStop}%
\bibitem [{\citenamefont {Wehinger}\ \emph {et~al.}(2018)\citenamefont
  {Wehinger}, \citenamefont {Fiolka}, \citenamefont {Lanza}, \citenamefont
  {Scatena}, \citenamefont {Kubus}, \citenamefont {Grockowiak}, \citenamefont
  {Coniglio}, \citenamefont {Graf}, \citenamefont {Skoulatos}, \citenamefont
  {Chen}, \citenamefont {Gukelberger}, \citenamefont {Casati}, \citenamefont
  {Zaharko}, \citenamefont {Macchi}, \citenamefont {Kr\"amer}, \citenamefont
  {Tozer}, \citenamefont {Mudry}, \citenamefont {Normand},\ and\ \citenamefont
  {R\"uegg}}]{Wehinger2016}%
  \BibitemOpen
  \bibfield  {author} {\bibinfo {author} {\bibfnamefont {B.}~\bibnamefont
  {Wehinger}}, \bibinfo {author} {\bibfnamefont {C.}~\bibnamefont {Fiolka}},
  \bibinfo {author} {\bibfnamefont {A.}~\bibnamefont {Lanza}}, \bibinfo
  {author} {\bibfnamefont {R.}~\bibnamefont {Scatena}}, \bibinfo {author}
  {\bibfnamefont {M.}~\bibnamefont {Kubus}}, \bibinfo {author} {\bibfnamefont
  {A.}~\bibnamefont {Grockowiak}}, \bibinfo {author} {\bibfnamefont {W.~A.}\
  \bibnamefont {Coniglio}}, \bibinfo {author} {\bibfnamefont {D.}~\bibnamefont
  {Graf}}, \bibinfo {author} {\bibfnamefont {M.}~\bibnamefont {Skoulatos}},
  \bibinfo {author} {\bibfnamefont {J.-H.}\ \bibnamefont {Chen}}, \bibinfo
  {author} {\bibfnamefont {J.}~\bibnamefont {Gukelberger}}, \bibinfo {author}
  {\bibfnamefont {N.}~\bibnamefont {Casati}}, \bibinfo {author} {\bibfnamefont
  {O.}~\bibnamefont {Zaharko}}, \bibinfo {author} {\bibfnamefont
  {P.}~\bibnamefont {Macchi}}, \bibinfo {author} {\bibfnamefont {K.~W.}\
  \bibnamefont {Kr\"amer}}, \bibinfo {author} {\bibfnamefont {S.}~\bibnamefont
  {Tozer}}, \bibinfo {author} {\bibfnamefont {C.}~\bibnamefont {Mudry}},
  \bibinfo {author} {\bibfnamefont {B.}~\bibnamefont {Normand}},\ and\ \bibinfo
  {author} {\bibfnamefont {C.}~\bibnamefont {R\"uegg}},\ }\bibfield  {title}
  {\bibinfo {title} {Giant pressure dependence and dimensionality switching in
  a metal-organic quantum antiferromagnet},\ }\href
  {https://doi.org/10.1103/PhysRevLett.121.117201} {\bibfield  {journal}
  {\bibinfo  {journal} {Phys. Rev. Lett.}\ }\textbf {\bibinfo {volume} {121}},\
  \bibinfo {pages} {117201} (\bibinfo {year} {2018})}\BibitemShut {NoStop}%
\bibitem [{\citenamefont {Woodward}\ \emph {et~al.}(2007)\citenamefont
  {Woodward}, \citenamefont {Gibson}, \citenamefont {Jameson}, \citenamefont
  {Landee}, \citenamefont {Turnbull},\ and\ \citenamefont
  {Willett}}]{Woodward2007}%
  \BibitemOpen
  \bibfield  {author} {\bibinfo {author} {\bibfnamefont {F.~M.}\ \bibnamefont
  {Woodward}}, \bibinfo {author} {\bibfnamefont {P.~J.}\ \bibnamefont
  {Gibson}}, \bibinfo {author} {\bibfnamefont {G.~B.}\ \bibnamefont {Jameson}},
  \bibinfo {author} {\bibfnamefont {C.~P.}\ \bibnamefont {Landee}}, \bibinfo
  {author} {\bibfnamefont {M.~M.}\ \bibnamefont {Turnbull}},\ and\ \bibinfo
  {author} {\bibfnamefont {R.~D.}\ \bibnamefont {Willett}},\ }\bibfield
  {title} {\bibinfo {title} {Two-dimensional {H}eisenberg antiferromagnets:
  Syntheses, x-ray structures, and magnetic behavior of
  [{C}u(pz)$_2$]({C}l{O}$_4$)$_2$, [{C}u(pz)$_2$]({B}{F}$_4$)$_2$, and
  [{C}u(pz)$_2$({N}{O}$_3$)]({P}{F}$_6$)},\ }\href
  {https://doi.org/10.1021/ic0621392} {\bibfield  {journal} {\bibinfo
  {journal} {Inorg. Chem.}\ }\textbf {\bibinfo {volume} {46}},\ \bibinfo
  {pages} {4256} (\bibinfo {year} {2007})}\BibitemShut {NoStop}%
\bibitem [{\citenamefont {Fischer}\ \emph {et~al.}(2000)\citenamefont
  {Fischer}, \citenamefont {Frey}, \citenamefont {Koch}, \citenamefont
  {K\"{o}nnecke}, \citenamefont {Pomjakushin}, \citenamefont {Schefer},
  \citenamefont {Thut}, \citenamefont {Schlumpf}, \citenamefont {B\"{u}rge},
  \citenamefont {Greuter}, \citenamefont {Bondt},\ and\ \citenamefont
  {Berruyer}}]{Fischer2000}%
  \BibitemOpen
  \bibfield  {author} {\bibinfo {author} {\bibfnamefont {P.}~\bibnamefont
  {Fischer}}, \bibinfo {author} {\bibfnamefont {G.}~\bibnamefont {Frey}},
  \bibinfo {author} {\bibfnamefont {M.}~\bibnamefont {Koch}}, \bibinfo {author}
  {\bibfnamefont {M.}~\bibnamefont {K\"{o}nnecke}}, \bibinfo {author}
  {\bibfnamefont {V.}~\bibnamefont {Pomjakushin}}, \bibinfo {author}
  {\bibfnamefont {J.}~\bibnamefont {Schefer}}, \bibinfo {author} {\bibfnamefont
  {R.}~\bibnamefont {Thut}}, \bibinfo {author} {\bibfnamefont {N.}~\bibnamefont
  {Schlumpf}}, \bibinfo {author} {\bibfnamefont {R.}~\bibnamefont {B\"{u}rge}},
  \bibinfo {author} {\bibfnamefont {U.}~\bibnamefont {Greuter}}, \bibinfo
  {author} {\bibfnamefont {S.}~\bibnamefont {Bondt}},\ and\ \bibinfo {author}
  {\bibfnamefont {E.}~\bibnamefont {Berruyer}},\ }\bibfield  {title} {\bibinfo
  {title} {High-resolution powder diffractometer hrpt for thermal neutrons at
  sinq},\ }\href {https://doi.org/10.1016/0168-9002(90)90141-R} {\bibfield
  {journal} {\bibinfo  {journal} {Physica B}\ }\textbf {\bibinfo {volume}
  {276}},\ \bibinfo {pages} {146} (\bibinfo {year} {2000})}\BibitemShut
  {NoStop}%
\bibitem [{\citenamefont {Rodr\'iguez-Carvajal}(1993)}]{Carvajal1993}%
  \BibitemOpen
  \bibfield  {author} {\bibinfo {author} {\bibfnamefont {J.}~\bibnamefont
  {Rodr\'iguez-Carvajal}},\ }\bibfield  {title} {\bibinfo {title} {Recent
  advances in magnetic structure determination by neutron powder diffraction},\
  }\href {https://doi.org/10.1016/S0921-4526(99)01399-X} {\bibfield  {journal}
  {\bibinfo  {journal} {Physica B}\ }\textbf {\bibinfo {volume} {192}},\
  \bibinfo {pages} {55} (\bibinfo {year} {1993})}\BibitemShut {NoStop}%
\bibitem [{\citenamefont {Clogston}\ \emph {et~al.}(1962)\citenamefont
  {Clogston}, \citenamefont {Gossard}, \citenamefont {Jaccarino},\ and\
  \citenamefont {Yafet}}]{Clogston1962}%
  \BibitemOpen
  \bibfield  {author} {\bibinfo {author} {\bibfnamefont {A.~M.}\ \bibnamefont
  {Clogston}}, \bibinfo {author} {\bibfnamefont {A.~C.}\ \bibnamefont
  {Gossard}}, \bibinfo {author} {\bibfnamefont {V.}~\bibnamefont {Jaccarino}},\
  and\ \bibinfo {author} {\bibfnamefont {Y.}~\bibnamefont {Yafet}},\ }\bibfield
   {title} {\bibinfo {title} {Orbital paramagnetism and the {K}night shift of
  $d$-band superconductors},\ }\href
  {https://doi.org/10.1103/PhysRevLett.9.262} {\bibfield  {journal} {\bibinfo
  {journal} {Phys. Rev. Lett.}\ }\textbf {\bibinfo {volume} {9}},\ \bibinfo
  {pages} {262} (\bibinfo {year} {1962})}\BibitemShut {NoStop}%
\bibitem [{\citenamefont {Shiroka}\ \emph {et~al.}(2015)\citenamefont
  {Shiroka}, \citenamefont {Pikulski}, \citenamefont {Zhigadlo}, \citenamefont
  {Batlogg}, \citenamefont {Mesot},\ and\ \citenamefont {Ott}}]{Shiroka2015}%
  \BibitemOpen
  \bibfield  {author} {\bibinfo {author} {\bibfnamefont {T.}~\bibnamefont
  {Shiroka}}, \bibinfo {author} {\bibfnamefont {M.}~\bibnamefont {Pikulski}},
  \bibinfo {author} {\bibfnamefont {N.~D.}\ \bibnamefont {Zhigadlo}}, \bibinfo
  {author} {\bibfnamefont {B.}~\bibnamefont {Batlogg}}, \bibinfo {author}
  {\bibfnamefont {J.}~\bibnamefont {Mesot}},\ and\ \bibinfo {author}
  {\bibfnamefont {H.-R.}\ \bibnamefont {Ott}},\ }\bibfield  {title} {\bibinfo
  {title} {Pairing of the weakly correlated electrons in the platinum-based
  centrosymmetric supeconductor {S}r{P}t$_3${P}},\ }\href
  {https://doi.org/10.1103/PhysRevB.91.245143} {\bibfield  {journal} {\bibinfo
  {journal} {Phys. Rev. B}\ }\textbf {\bibinfo {volume} {91}},\ \bibinfo
  {pages} {245143} (\bibinfo {year} {2015})}\BibitemShut {NoStop}%
\bibitem [{\citenamefont {Bryce}\ and\ \citenamefont
  {Sward}(2006)}]{Bryce2006}%
  \BibitemOpen
  \bibfield  {author} {\bibinfo {author} {\bibfnamefont {D.~L.}\ \bibnamefont
  {Bryce}}\ and\ \bibinfo {author} {\bibfnamefont {G.~D.}\ \bibnamefont
  {Sward}},\ }\bibfield  {title} {\bibinfo {title} {Solid-state {NMR}
  spectroscopy of the quadrupolar halogens: chlorine-35/37, bromine-79/81, and
  iodine-127},\ }\href {https://doi.org/10.1002/mrc.1741} {\bibfield  {journal}
  {\bibinfo  {journal} {Magn. Reson. Chem.}\ }\textbf {\bibinfo {volume}
  {44}},\ \bibinfo {pages} {409} (\bibinfo {year} {2006})}\BibitemShut
  {NoStop}%
\bibitem [{\citenamefont {Borsa}(2007)}]{Borsa2007}%
  \BibitemOpen
  \bibfield  {author} {\bibinfo {author} {\bibfnamefont {F.}~\bibnamefont
  {Borsa}},\ }\bibinfo {title} {Phase transitions and critical phenomena in
  solids},\ in\ \href {https://doi.org/10.1002/9780470034590.emrstm0390} {\emph
  {\bibinfo {booktitle} {eMagRes}}},\ \bibinfo {editor} {edited by\ \bibinfo
  {editor} {\bibfnamefont {R.~K.}\ \bibnamefont {Harris}}\ and\ \bibinfo
  {editor} {\bibfnamefont {R.~E.}\ \bibnamefont {Wasylishen}}}\ (\bibinfo
  {publisher} {John Wiley \& Sons},\ \bibinfo {address} {Hoboken, NJ},\
  \bibinfo {year} {2007})\BibitemShut {NoStop}%
\bibitem [{Note1()}]{Note1}%
  \BibitemOpen
  \bibinfo {note} {The sensitivity of the NMR relaxation rate to phase
  transitions arises from the so-called \protect \emph {critical slowing down}
  of fluctuations in the proximity of a transition temperature $T_{c}$. The
  presence of long-range spatial correlations close to $T_{c}$ implies also
  slow time fluctuations. The latter, who match well the nuclear-spin time
  scales, provide a very effective relaxation channel and give rise to a peak
  in $1/T_{1}(T)$.}\BibitemShut {Stop}%
\bibitem [{\citenamefont {McDowell}(1995)}]{Mcdowell1995}%
  \BibitemOpen
  \bibfield  {author} {\bibinfo {author} {\bibfnamefont {A.~F.}\ \bibnamefont
  {McDowell}},\ }\bibfield  {title} {\bibinfo {title} {Magnetization-recovery
  curves for quadrupolar spins},\ }\href
  {https://doi.org/10.1006/jmra.1995.1087} {\bibfield  {journal} {\bibinfo
  {journal} {J. Magn. Reson. Ser. A}\ }\textbf {\bibinfo {volume} {113}},\
  \bibinfo {pages} {242} (\bibinfo {year} {1995})}\BibitemShut {NoStop}%
\bibitem [{\citenamefont {Schettino}\ \emph {et~al.}(1972)\citenamefont
  {Schettino}, \citenamefont {Sbrana},\ and\ \citenamefont
  {Righini}}]{Schettino1972}%
  \BibitemOpen
  \bibfield  {author} {\bibinfo {author} {\bibfnamefont {V.}~\bibnamefont
  {Schettino}}, \bibinfo {author} {\bibfnamefont {G.}~\bibnamefont {Sbrana}},\
  and\ \bibinfo {author} {\bibfnamefont {R.}~\bibnamefont {Righini}},\
  }\bibfield  {title} {\bibinfo {title} {Evidence for a phase transition in
  crystalline pyrazine},\ }\href {https://doi.org/10.1016/0009-2614(72)85063-2}
  {\bibfield  {journal} {\bibinfo  {journal} {Chem. Phys. Lett.}\ }\textbf
  {\bibinfo {volume} {13}},\ \bibinfo {pages} {284} (\bibinfo {year}
  {1972})}\BibitemShut {NoStop}%
\bibitem [{\citenamefont {Boyd}\ \emph {et~al.}(1979)\citenamefont {Boyd},
  \citenamefont {Comper},\ and\ \citenamefont {Ferguson}}]{Boyd1979}%
  \BibitemOpen
  \bibfield  {author} {\bibinfo {author} {\bibfnamefont {R.~K.}\ \bibnamefont
  {Boyd}}, \bibinfo {author} {\bibfnamefont {J.}~\bibnamefont {Comper}},\ and\
  \bibinfo {author} {\bibfnamefont {G.}~\bibnamefont {Ferguson}},\ }\bibfield
  {title} {\bibinfo {title} {Entropy changes and structural implications for
  crystalline phases of pyrazine},\ }\href {https://doi.org/10.1139/v79-498}
  {\bibfield  {journal} {\bibinfo  {journal} {Can. J. Chem.}\ }\textbf
  {\bibinfo {volume} {57}},\ \bibinfo {pages} {3056} (\bibinfo {year}
  {1979})}\BibitemShut {NoStop}%
\bibitem [{\citenamefont {Hammar}\ \emph {et~al.}(1999)\citenamefont {Hammar},
  \citenamefont {Stone}, \citenamefont {Reich}, \citenamefont {Broholm},
  \citenamefont {Gibson}, \citenamefont {Turnbull}, \citenamefont {Landee},\
  and\ \citenamefont {Oshikawa}}]{Hammar1999}%
  \BibitemOpen
  \bibfield  {author} {\bibinfo {author} {\bibfnamefont {P.~R.}\ \bibnamefont
  {Hammar}}, \bibinfo {author} {\bibfnamefont {M.~B.}\ \bibnamefont {Stone}},
  \bibinfo {author} {\bibfnamefont {D.~H.}\ \bibnamefont {Reich}}, \bibinfo
  {author} {\bibfnamefont {C.}~\bibnamefont {Broholm}}, \bibinfo {author}
  {\bibfnamefont {P.~J.}\ \bibnamefont {Gibson}}, \bibinfo {author}
  {\bibfnamefont {M.~M.}\ \bibnamefont {Turnbull}}, \bibinfo {author}
  {\bibfnamefont {C.~P.}\ \bibnamefont {Landee}},\ and\ \bibinfo {author}
  {\bibfnamefont {M.}~\bibnamefont {Oshikawa}},\ }\bibfield  {title} {\bibinfo
  {title} {Characterization of a quasi-one-dimensional spin-1/2 magnet which is
  gapless and paramagnetic for $g \mu_\mathrm{B} {H} \lesssim {J}$ and
  ${k}_\mathrm{B}{T} \ll {J}$},\ }\href
  {https://doi.org/10.1103/PhysRevB.59.1008} {\bibfield  {journal} {\bibinfo
  {journal} {Phys. Rev. B}\ }\textbf {\bibinfo {volume} {59}},\ \bibinfo
  {pages} {1008} (\bibinfo {year} {1999})}\BibitemShut {NoStop}%
\bibitem [{\citenamefont {O'Neal}\ \emph {et~al.}(2014)\citenamefont {O'Neal},
  \citenamefont {Brinzari}, \citenamefont {Wright}, \citenamefont {Ma},
  \citenamefont {Giri}, \citenamefont {Schlueter}, \citenamefont {Wang},
  \citenamefont {Jena}, \citenamefont {Liu},\ and\ \citenamefont
  {Musfeldt}}]{Oneal2014}%
  \BibitemOpen
  \bibfield  {author} {\bibinfo {author} {\bibfnamefont {K.~R.}\ \bibnamefont
  {O'Neal}}, \bibinfo {author} {\bibfnamefont {T.~V.}\ \bibnamefont
  {Brinzari}}, \bibinfo {author} {\bibfnamefont {J.~B.}\ \bibnamefont
  {Wright}}, \bibinfo {author} {\bibfnamefont {C.}~\bibnamefont {Ma}}, \bibinfo
  {author} {\bibfnamefont {S.}~\bibnamefont {Giri}}, \bibinfo {author}
  {\bibfnamefont {J.~A.}\ \bibnamefont {Schlueter}}, \bibinfo {author}
  {\bibfnamefont {Q.}~\bibnamefont {Wang}}, \bibinfo {author} {\bibfnamefont
  {P.}~\bibnamefont {Jena}}, \bibinfo {author} {\bibfnamefont {Z.}~\bibnamefont
  {Liu}},\ and\ \bibinfo {author} {\bibfnamefont {J.~L.}\ \bibnamefont
  {Musfeldt}},\ }\bibfield  {title} {\bibinfo {title} {Pressure-induced
  magnetic crossover driven by hydrogen bonding in
  {C}u{F}$_2$({H}$_2${O})$_2$(3-chloropyridine)},\ }\href
  {https://doi.org/10.1038/srep06054} {\bibfield  {journal} {\bibinfo
  {journal} {Sci. Rep.}\ }\textbf {\bibinfo {volume} {4}},\ \bibinfo {pages}
  {6054} (\bibinfo {year} {2014})}\BibitemShut {NoStop}%
\bibitem [{\citenamefont {Le~Bail}(2005)}]{Lebail2005}%
  \BibitemOpen
  \bibfield  {author} {\bibinfo {author} {\bibfnamefont {A.}~\bibnamefont
  {Le~Bail}},\ }\bibfield  {title} {\bibinfo {title} {Whole powder pattern
  decomposition methods and applications: A retrospection},\ }\href
  {https://doi.org/10.1154/1.2135315} {\bibfield  {journal} {\bibinfo
  {journal} {Powder Diffr.}\ }\textbf {\bibinfo {volume} {20}},\ \bibinfo
  {pages} {316} (\bibinfo {year} {2005})}\BibitemShut {NoStop}%
\bibitem [{\citenamefont {Medarde}\ \emph {et~al.}(2013)\citenamefont
  {Medarde}, \citenamefont {Mena}, \citenamefont {Gavilano}, \citenamefont
  {Pomjakushina}, \citenamefont {Sugiyama}, \citenamefont {Kamazawa},
  \citenamefont {Pomjakushin}, \citenamefont {Sheptyakov}, \citenamefont
  {Batlogg}, \citenamefont {Ott}, \citenamefont {Mansson},\ and\ \citenamefont
  {Juranyi}}]{Medarde2013}%
  \BibitemOpen
  \bibfield  {author} {\bibinfo {author} {\bibfnamefont {M.}~\bibnamefont
  {Medarde}}, \bibinfo {author} {\bibfnamefont {M.}~\bibnamefont {Mena}},
  \bibinfo {author} {\bibfnamefont {J.~L.}\ \bibnamefont {Gavilano}}, \bibinfo
  {author} {\bibfnamefont {E.}~\bibnamefont {Pomjakushina}}, \bibinfo {author}
  {\bibfnamefont {J.}~\bibnamefont {Sugiyama}}, \bibinfo {author}
  {\bibfnamefont {K.}~\bibnamefont {Kamazawa}}, \bibinfo {author}
  {\bibfnamefont {V.}~\bibnamefont {Pomjakushin}}, \bibinfo {author}
  {\bibfnamefont {D.}~\bibnamefont {Sheptyakov}}, \bibinfo {author}
  {\bibfnamefont {B.}~\bibnamefont {Batlogg}}, \bibinfo {author} {\bibfnamefont
  {H.-R.}\ \bibnamefont {Ott}}, \bibinfo {author} {\bibfnamefont
  {M.}~\bibnamefont {Mansson}},\ and\ \bibinfo {author} {\bibfnamefont
  {F.}~\bibnamefont {Juranyi}},\ }\bibfield  {title} {\bibinfo {title} {{1D} to
  {2D} {Na}$^+$ ion diffusion inherently linked to structural transitions in
  {Na}$_{0.7}${CoO}$_2$},\ }\href
  {https://doi.org/10.1103/PhysRevLett.110.266401} {\bibfield  {journal}
  {\bibinfo  {journal} {Phys. Rev. Lett.}\ }\textbf {\bibinfo {volume} {110}},\
  \bibinfo {pages} {266401} (\bibinfo {year} {2013})}\BibitemShut {NoStop}%
\bibitem [{\citenamefont {Morin}\ \emph {et~al.}(2016)\citenamefont {Morin},
  \citenamefont {Canevet}, \citenamefont {Raynaud}, \citenamefont {Bartkowiak},
  \citenamefont {Sheptyakov}, \citenamefont {Ban}, \citenamefont {Kenzelmann},
  \citenamefont {Pomjakushina}, \citenamefont {Conder},\ and\ \citenamefont
  {Medarde}}]{Morin2016}%
  \BibitemOpen
  \bibfield  {author} {\bibinfo {author} {\bibfnamefont {M.}~\bibnamefont
  {Morin}}, \bibinfo {author} {\bibfnamefont {E.}~\bibnamefont {Canevet}},
  \bibinfo {author} {\bibfnamefont {A.}~\bibnamefont {Raynaud}}, \bibinfo
  {author} {\bibfnamefont {M.}~\bibnamefont {Bartkowiak}}, \bibinfo {author}
  {\bibfnamefont {D.}~\bibnamefont {Sheptyakov}}, \bibinfo {author}
  {\bibfnamefont {V.}~\bibnamefont {Ban}}, \bibinfo {author} {\bibfnamefont
  {M.}~\bibnamefont {Kenzelmann}}, \bibinfo {author} {\bibfnamefont
  {E.}~\bibnamefont {Pomjakushina}}, \bibinfo {author} {\bibfnamefont
  {K.}~\bibnamefont {Conder}},\ and\ \bibinfo {author} {\bibfnamefont
  {M.}~\bibnamefont {Medarde}},\ }\bibfield  {title} {\bibinfo {title} {Tuning
  magnetic spirals beyond room temperature with chemical disorder},\ }\href
  {https://doi.org/10.1038/ncomms13758} {\bibfield  {journal} {\bibinfo
  {journal} {Nat. Commun.}\ }\textbf {\bibinfo {volume} {7}},\ \bibinfo {pages}
  {13758} (\bibinfo {year} {2016})}\BibitemShut {NoStop}%
\bibitem [{\citenamefont {Shang}\ \emph {et~al.}(2018)\citenamefont {Shang},
  \citenamefont {Canevet}, \citenamefont {Morin}, \citenamefont {Sheptyakov},
  \citenamefont {Fernandez-Diaz}, \citenamefont {Pomjakushina},\ and\
  \citenamefont {Medarde}}]{Shang2018}%
  \BibitemOpen
  \bibfield  {author} {\bibinfo {author} {\bibfnamefont {T.}~\bibnamefont
  {Shang}}, \bibinfo {author} {\bibfnamefont {E.}~\bibnamefont {Canevet}},
  \bibinfo {author} {\bibfnamefont {M.}~\bibnamefont {Morin}}, \bibinfo
  {author} {\bibfnamefont {D.}~\bibnamefont {Sheptyakov}}, \bibinfo {author}
  {\bibfnamefont {M.}~\bibnamefont {Fernandez-Diaz}}, \bibinfo {author}
  {\bibfnamefont {E.}~\bibnamefont {Pomjakushina}},\ and\ \bibinfo {author}
  {\bibfnamefont {M.}~\bibnamefont {Medarde}},\ }\bibfield  {title} {\bibinfo
  {title} {Design of magnetic spirals in layered perovskites: Extending the
  stability range far beyond room temperature},\ }\href
  {https://doi.org/10.1126/sciadv.aau6386} {\bibfield  {journal} {\bibinfo
  {journal} {Sci. Adv.}\ }\textbf {\bibinfo {volume} {4}},\ \bibinfo {pages}
  {eaau6386} (\bibinfo {year} {2018})}\BibitemShut {NoStop}%
\bibitem [{Note2()}]{Note2}%
  \BibitemOpen
  \bibinfo {note} {We recall that ADPs, which describe the anisotropic thermal
  motion through a symmetric rank-2 tensor, consist of six independent
  parameters. In the isotropic case, the off-diagonal terms clearly vanish.
  Most often $U_{ij}$ are used to express ADPs, since they represent directly
  the mean-square atomic displacements (some values relevant to our case are
  reported in Table~\ref {tab:structure}). An alternative, closely related
  quantity is $B_{ij} = 8\pi ^2 U_{ij}$. Computer programs generally output the
  $\beta _{ij}$ values, defined as $\beta _{ij} = B_{ij}x^{\star }_ix^{\star
  }_j/4$, since these minimize computation time and can be related directly to
  the reciprocal-cell parameters $x^{\star }_i$ and $x^{\star }_j$. The
  diagonal ADPs describe displacements along three mutually perpendicular axes
  of the ellipsoid and, hence, are always positive. On the other hand, since
  the other elements of the ADP tensor establish the orientation of the
  ellipsoid with respect to the crystal-lattice coordinate system, the
  off-diagonal elements can be either positive or negative, under the
  structural constrain $\beta _{ii}\beta _{jj} > \beta ^2_{ij}$, which is
  confirmed from our data and is a necessary requirement for the physical
  validity of the refinement.}\BibitemShut {Stop}%
\bibitem [{Note3()}]{Note3}%
  \BibitemOpen
  \bibinfo {note} {The quantities marked with (*) in Table I were constrained
  to be identical for the same atomic species. The occupation of the D sites
  was found to be very close to 100\%. The occupation of the remaining atomic
  sites was 100\%}\BibitemShut {NoStop}%
\bibitem [{Note4()}]{Note4}%
  \BibitemOpen
  \bibinfo {note} {The Supplemental Material available at [URL] includes the
  two structural .cif files used for refining the neutron-diffraction patterns
  at 295 K, by assuming as valid the space groups $C2$ and $C2/m$,
  respectively.}\BibitemShut {Stop}%
\end{thebibliography}

%

\end{document}